\documentclass[journal]{IEEEtran}

\usepackage{xcolor,soul,framed} 

\colorlet{shadecolor}{yellow}
\usepackage[pdftex]{graphicx}
\graphicspath{{../pdf/}{../jpeg/}}
\DeclareGraphicsExtensions{.pdf,.jpeg,.png}

\usepackage[cmex10]{amsmath}
\usepackage{array}
\usepackage{booktabs}
\usepackage{tabularx}
\usepackage{mdwmath}
\usepackage{mdwtab}
\usepackage{eqparbox}
\usepackage{url}
\usepackage{algorithm}
\usepackage{algorithmicx}
\usepackage{algcompatible}
\usepackage{mathrsfs}
\usepackage{amssymb} 
\usepackage{amsmath}
\usepackage{mathtools}
\usepackage{mathrsfs}
\usepackage{dutchcal}
\usepackage{url}
\usepackage{multirow}
\usepackage{breakurl}
\usepackage{enumerate}
\usepackage[hidelinks]{hyperref}

\begin{document}
	\bstctlcite{IEEEexample:BSTcontrol}
	\title{A Lightweight Privacy-Preserving Smart Metering Billing Protocol with Dynamic Tariff Policy Adjustment}
	\author{Farid~Zaredar, and 
		Morteza~Amini 
	
	\thanks{F. Zaredar is with Department of Computer Engineering, Sharif University of Technology, Tehran, Iran (e-mail: farid.zaredar78@sharif.edu).}
	
	\thanks{M. Amini is with Department of Computer Engineering, Sharif University of Technology, Tehran, Iran (e-mail: amini@sharif.edu).}
}


	\maketitle

\begin{abstract}
	Fine-grained smart-meter consumption data are essential for accurate real-time-tariff billing, but their collection can reveal sensitive information about customers' daily activities and consumption patterns. To address this privacy challenge, we propose a lightweight privacy-preserving smart metering protocol for real-time tariff billing with dynamic tariff policy adjustment. Our scheme employs an efficient data perturbation mechanism with a tariff-weighted zero-sum property, allowing the utility provider to compute the exact customer bill from perturbed
	fine-grained consumption readings. 	The protocol further supports authorized tariff policy adjustments after the consumption readings have already been reported. For proportional (Type I)
	adjustments, all previously reported noisy readings are reused without any additional smart-meter report. For non-proportional (Type II) adjustments, the first $L-1$ noisy readings (where L is the number of intervals) are reused and only the final corrected noisy consumption value is updated, thereby reducing the additional communication, computation, and storage resources required for tariff-policy adjustment. The
	number of accepted Type II adjustments is also bounded to limit the additional exact algebraic information introduced by repeated non-proportional tariff changes. We evaluate the proposed scheme in terms of computational, memory, communication, and privacy characteristics. The protocol requires approximately $3.94540$ seconds of execution time for a complete year, and the experimental results confirm exact billing to numerical precision. Privacy is evaluated through statistical characterization, reconstruction-based analysis, tariff-adjustment analysis, and Jensen--Shannon divergence. The results show that increasing the evaluated noise scale reduces the additional target-specific reconstruction value available to the considered attack while increasing the distributional
	difference between the original and perturbed consumption data.
\end{abstract}


	%
	\IEEEpeerreviewmaketitle

	\section{Introduction} 
	\IEEEPARstart{T}{he} continuous growth of smart cities, has led to significant changes in energy demand and supply patterns. Traditional electrical grids can no longer adequately support the evolving energy requirements of diverse sectors in modern urban environments. Consequently, the concept of smart grids has been introduced to address energy infrastructure needs, providing capabilities such as dynamic billing, real-time monitoring, efficient grid management, rapid self-healing, and accurate load forecasting \cite{souri2014smart}. 

	Today, Smart grids are expanding rapidly across the globe. For instance, the United States invested \$6.4 billion on smart grid technology in 2018 and this investment is projected to grow to \$16.4 billion annually by 2026 \cite{DOE2020SmartGrid}. The smart grid market in the United States witnesses significant growth and is forecasted to reach \$64.34 billion by 2030 \cite{MarketsAndData2025}. Similarly, According to the IMARC group \cite{IMARC2025}, the smart grid market in the United Kingdom was valued at \$1.84 billion in 2023 and is predicted to reach \$8.46 billion by 2032, reflecting a compound annual growth rate (CAGR) of 16.30\% during the period 2024 to 2032. The European Union (EU) has adopted a rigorous strategy for deploying smart metering systems, investing €47 billion in this sector by 2030 \cite{EuropeanCommission2025}.
	
	Smart grids are more reliable, efficient, and sustainable compared to traditional power grids \cite{kumar2019smart}. Unlike the earlier technology, automated meter reading (AMR), which enabled only unidirectional communication \cite{siddiqui2012smart}, the advanced metering infrastructure (AMI) facilitates bidirectional communication between smart meters and the control center. This capability allows intelligent meters to transmit consumption data frequently (e.g., every 15 or 30 minutes) and at fine granularity to the meter data management system (MDMS). The MDMS collects and processes consumers' consumption readings to support various functional utilities, such as billing, operational, and value-added services \cite{asghar2017smart, kayalvizhy2021survey, ansari2022state}. Furthermore, these modern power grids can significantly reduce carbon emissions by integrating with renewable energy sources (RES) and utilizing clean energy sources, such as solar and wind power \cite{kua2023privacy}. 
	
	While smart grids offer numerous valuable services to both customers and utility providers, they also raise notable privacy concerns. For instance, the collection of fine-grained consumption data for billing, can unintentionally reveal users' daily routines, such as leaving for work (i.e., individual absence), cooking, watching TV, working with home computers or smart devices, and sleeping patterns, thereby compromising inhabitants' privacy. As a consequence, safeguarding customer privacy has become a critical issue that must be addressed.  
	
	In the literature, numerous privacy-preserving smart metering protocols have been proposed. These schemes preserve consumers' privacy while enabling essential utilities, including bill generation, load monitoring, and anomaly detection (e.g., outage diagnosis). Prior studies have employed various privacy-enhancing techniques, such as secure multi-party computation, zero-knowledge proofs, commitment schemes, homomorphic encryptions, and data masking. However, these approaches often introduce significant computational, memory, and communication overhead. 
	
	To mitigate such performance challenges, we propose a lightweight privacy-preserving smart metering protocol designed to enable real-time tariff billing with dynamic tariff policy adjustments. Our protocol leverages an efficient data perturbation technique to obscure fine-grained consumption readings and reduce the disclosure of users' detailed energy usage information. Additionally, our scheme is adaptable to tariff policy changes adopted by energy suppliers. Unlike the representative schemes reviewed in Section II, our protocol explicitly supports tariff policy adjustments after consumption readings have already been reported while efficiently reusing the previously reported noisy readings. Since IoT devices such as smart meters have limited computational resources, our approach employs a resource-friendly data perturbation technique to introduce noise into consumption values.
	
	The main contributions of this work are summarized as follows:
	\begin{itemize}
		
		\item \textbf{Lightweight privacy-aware real-time-tariff billing:}
		We introduce a lightweight data perturbation mechanism that protects
		fine-grained consumption readings while preserving exact billing. By enforcing
		a tariff-weighted zero-sum property on the generated noise values, the utility
		provider can compute the customer's exact bill directly from the reported
		perturbed consumption values.
		
		\item \textbf{Efficient dynamic tariff policy adjustment:}
		We support authorized tariff policy changes after the consumption readings
		have already been reported. For proportional Type I adjustments, all
		previously reported noisy readings are reused without any additional report
		from the smart meter. For non-proportional Type II adjustments, the first
		$L-1$ reports are reused and only the final corrected noisy consumption value
		is updated. Reusing the previously reported readings avoids regenerating and
		retransmitting an entire consumption vector, thereby reducing the additional
		communication bandwidth, computation, and storage resources required for
		tariff-policy adjustment.
	
		\item \textbf{Controlled disclosure under repeated tariff adjustments:}
		We bound the number of accepted non-proportional tariff adjustments by
		$K_{\max}$ and analytically characterize the additional exact billing
		information introduced by repeated Type II adjustments. This provides an
		explicit mechanism for limiting the incremental algebraic disclosure caused
		by repeated tariff-policy changes.
		
		\item \textbf{Comprehensive performance and privacy evaluation:}
		We evaluate the proposed protocol in terms of computational, memory, and
		communication overhead, together with statistical, reconstruction-based, and
		distributional privacy analyses. The experimental results confirm exact
		billing to numerical precision while demonstrating lightweight operation and
		the privacy--utility behavior of the perturbation mechanism.
		
	\end{itemize}
	
	The rest of the paper is organized as follows. Section II reviews related
	work. Section III presents the system and threat models and design goals.
	Section IV describes the proposed protocol, Section V presents its performance
	and privacy evaluation, and Section VI concludes the paper.
	
	\section{Related Work}
	
	Numerous privacy-preserving smart metering schemes have been proposed to
	support electricity billing services. Depending on the adopted tariff model,
	billing may require coarse-grained or fine-grained consumption information
	\cite{sultan2019privacy}. However, attributable fine-grained readings can
	reveal sensitive information about customers, including daily activities and
	consumption patterns. Consequently, privacy-preserving billing protocols aim
	to enable accurate bill computation while limiting the disclosure of
	fine-grained consumption data.
	
	For the purpose of this work, we organize the most relevant approaches into
	three categories according to the entity responsible for the privacy-aware
	billing process: (1) customer-side secure billing, (2) intermediary-assisted
	secure billing, and (3) utility-centric secure billing.

	\subsection{Customer-Side Secure Billing}
	
	In these approaches, the billing computation is performed at the customer's
	premises, typically by the smart meter or an additional customer-side device.
	Trusted hardware, commitment schemes, and zero-knowledge proofs are commonly
	employed to prevent the disclosure of fine-grained consumption readings while
	allowing the energy supplier to verify the resulting bill.
	
	Jawurek et al. \cite{jawurek2011plug} proposed a privacy-preserving protocol
	based on a plug-in privacy component that computes the customer's bill, while
	Pedersen commitments enable the supplier to verify its correctness. Similarly,
	Molina-Markham et al. \cite{molina2010private} allow smart meters to compute
	customers' bills locally and employ anonymization and zero-knowledge
	techniques for verification.
	
	Rial et al. \cite{rial2011privacy} employ zero-knowledge proofs together with
	Groth integer commitments \cite{groth2005non} to provide secure billing at the
	customer's premises. Their subsequent work \cite{rial2018privacy} extends the
	model to multi-user and multi-meter settings and supports complex tariff
	policies. Romdhane et al. \cite{ben2021efficient} similarly employ
	elliptic-curve Pedersen commitments for private time-of-use billing.
	
	Other approaches rely more directly on trusted hardware. Eccles et al.
	\cite{eccles2017performance} presented time-of-use billing protocols using
	trusted, tamper-resistant smart meters, while Petrlic
	\cite{petrlic2010privacy} incorporates trusted platform modules and remote
	attestation to verify customer-side billing software. Zhang et al.
	\cite{zhang2020privacy} designed a multi-channel architecture in which smart
	meters compute bills locally and employ zero-knowledge proofs to demonstrate
	correctness. Overall, these approaches limit the utility provider's access to
	detailed readings, but shift billing computation, trusted execution, or
	cryptographic verification toward customer-side devices.

	\subsection{Intermediary-Assisted Secure Billing}
	
	In this class, the billing process is performed or assisted by intermediary
	entities in a privacy-preserving manner. Such entities can reduce the
	processing burden on resource-constrained smart meters, but may introduce
	additional trust assumptions and infrastructure requirements.
	
	Bohli et al. \cite{bohli2010privacy} proposed a privacy model in which a
	trusted third party collects metering data and performs temporal aggregation
	before reporting the total consumption of each meter to the energy supplier.
	In contrast, Wang et al. \cite{wang2023privacy} employ a fog-based
	architecture in which fog nodes aggregate encrypted readings and additional
	entities support real-time pricing and dynamic billing.
	
	Braeken et al. \cite{braeken2018efficient} proposed a privacy-preserving
	aggregation and dynamic billing scheme involving smart meters, data
	concentrators, and an operation center. The operation center determines
	fluctuating electricity prices, while protected metering reports are processed
	through the data concentrator to compute customer bills.
	
	Li et al. \cite{li2023fine} similarly employ fog servers and an outsourcing
	service provider to process encrypted fine-grained readings and support
	dynamic electricity pricing. Xu et al. \cite{xu2023privacy}, in contrast,
	provide a flexible framework based on homomorphic encryption and secure
	multi-party computation under different trust assumptions.
	
	Wagh et al. \cite{wagh2024distributed} proposed a distributed
	privacy-preserving framework supporting time-of-use billing and metering-data
	integrity verification under a malicious adversarial setting.
	
	More recent approaches employ blockchain-based infrastructures. Oksuz
	\cite{oksuz2024consortium} combines data randomization and pseudonymous
	identities with a consortium blockchain to support data aggregation, dynamic
	pricing, and dynamic billing. Kayumov et al. \cite{kayumov2025trusted}
	similarly employ a permissioned blockchain and smart contracts, while
	privacy-sensitive billing operations are executed inside a trusted execution
	environment.

	\subsection{Utility-Centric Secure Billing}
	
	In this category, the utility provider or system operator ultimately computes
	customers' bills using protected or transformed metering information.
	
	Efthymiou et al. \cite{efthymiou2010smart} proposed a privacy-aware smart
	metering protocol based on separate low-frequency and high-frequency
	identifiers. Low-frequency attributable readings are used for billing, while
	high-frequency readings employ different identifiers for operational
	services. Ababneh et al. \cite{ababneh2022private}, in contrast, protect
	high-frequency readings through blinding and Shamir secret sharing before the
	utility obtains the information required for bill computation.
	
	The PPETD framework \cite{nabil2019ppetd} employs secret sharing to mask
	periodically reported consumption readings. The masks cancel during
	aggregation, enabling billing and load monitoring without revealing
	individual fine-grained readings to the system operator. Ibrahem et al.
	\cite{ibrahem2021efficient} employ functional encryption, allowing the system
	operator to compute customers' bills according to dynamic pricing over
	encrypted fine-grained readings.
	
	Abidin et al. \cite{abidin2018secure} proposed a particularly relevant
	cancellation-based billing mechanism. Each smart meter locally computes its
	billing value for each period and masks it using a random value, where the
	random masks over the billing period sum to zero. The supplier can therefore
	recover the final monthly bill without learning the individual per-period
	billing values. Similar to our approach, this scheme exploits cancellation to
	preserve billing correctness. However, the masking is applied to locally
	computed billing values, whereas our scheme perturbs fine-grained consumption
	reports using a tariff-weighted correction and enables the utility provider to
	perform the billing computation.
	
	Overall, several representative approaches already support flexible,
	time-varying, or real-time pricing. However, we did not identify explicit
	support in these works for recomputing a bill after an authorized tariff
	policy revision by efficiently reusing the previously reported
	privacy-protected consumption data. Table~\ref{table:billing-approaches}
	focuses on the schemes most closely related to this design objective.

	\begin{table*}[!t]
		\centering
		\caption{Comparison of representative privacy-preserving smart metering
			billing schemes.}
		\label{table:billing-approaches}
		\renewcommand{\arraystretch}{1.25}
		\setlength{\tabcolsep}{5pt}
		\footnotesize
		
		\begin{tabularx}{\textwidth}{
				@{}
				>{\raggedright\arraybackslash}p{2.55cm}
				>{\raggedright\arraybackslash}X
				>{\raggedright\arraybackslash}X
				>{\centering\arraybackslash}p{1.25cm}
				>{\centering\arraybackslash}p{2.25cm}
				>{\centering\arraybackslash}p{2.60cm}
				@{}}
			
			\toprule
			
			\multirow{2}{*}{\textbf{Scheme}} &
			\multirow{2}{*}{\textbf{Privacy mechanism}} &
			\multirow{2}{*}{\shortstack[l]{\textbf{Main SM-side}\\\textbf{operation}}} &
			\multirow{2}{*}{\textbf{Pricing}} &
			\multicolumn{2}{c}{\textbf{Tariff-adjustment capability}} \\
			
			\cmidrule(lr){5-6}
			
			&
			&
			&
			&
			\shortstack{\textbf{Tariff revision}\\\textbf{after reporting}} &
			\shortstack{\textbf{Previously reported}\\\textbf{readings reused}} \\
			
			\midrule
			
			Rial et al.
			\cite{rial2011privacy,rial2018privacy}
			&
			Commitments + ZKP
			&
			Digital signatures
			&
			Flexible
			&
			NR
			&
			NR
			\\
			
			\addlinespace[1pt]
			
			Abidin et al.
			\cite{abidin2018secure}
			&
			Zero-sum masking
			&
			PRNG + arithmetic
			&
			TOU
			&
			NR
			&
			NR
			\\
			
			\addlinespace[1pt]
			
			Braeken et al.
			\cite{braeken2018efficient}
			&
			Secure aggregation
			&
			ECC + sym. enc. + hash
			&
			Dynamic
			&
			NR
			&
			NR
			\\
			
			\addlinespace[1pt]
			
			Ibrahem et al.
			\cite{ibrahem2021efficient}
			&
			Functional encryption
			&
			FE encryption
			&
			Dynamic
			&
			NR
			&
			NR
			\\
			
			\addlinespace[1pt]
			
			Wang et al.
			\cite{wang2023privacy}
			&
			Homomorphic encryption
			&
			ElGamal encryption
			&
			RTP
			&
			NR
			&
			NR
			\\
			
			\addlinespace[1pt]
			
			Wagh et al.
			\cite{wagh2024distributed}
			&
			Distributed aggregation
			&
			Secret sharing + commitments
			&
			TOU
			&
			NR
			&
			NR
			\\
			
			\addlinespace[1pt]
			
			Oksuz
			\cite{oksuz2024consortium}
			&
			Randomization + blockchain
			&
			Randomization + auth.
			&
			Dynamic
			&
			NR
			&
			NR
			\\
			
			\midrule[0.9pt]
			
			\textbf{Our scheme}
			&
			\textbf{Tariff-weighted perturbation}
			&
			\textbf{PRNG + arithmetic}
			&
			\textbf{RTP}
			&
			\textbf{Type I / Type II}
			&
			\shortstack{\textbf{I: all}\\\textbf{II: $L-1$}}
			\\
			
			\bottomrule
		\end{tabularx}
		
		\vspace{1mm}
		\raggedright
		\scriptsize
		\textit{Note:} SM: smart meter; ZKP: zero-knowledge proof;
		ECC: elliptic-curve cryptography; sym. enc.: symmetric encryption;
		FE: functional encryption; TOU: time-of-use pricing;
		RTP: real-time pricing; auth.: authentication.
		NR indicates that explicit support for tariff-policy revision after the
		consumption readings have already been reported, or reuse of those reports
		following such a revision, was not identified in the cited scheme.
		For our scheme, Type I reuses all previously reported noisy readings,
		whereas Type II reuses the first $L-1$ readings and updates only the final
		noisy reading.
	\end{table*}

	\section{Problem Statement and Design Goals}
	
	Although attributable fine-grained consumption readings are required for various smart-grid services, they may reveal sensitive information about customers' activities and consumption patterns. Moreover, many existing privacy-enhancing approaches introduce significant computational and algorithmic complexity, thereby demanding considerable resources from both smart meters and control centers. To address these privacy and resource-constrained concerns, we propose a lightweight privacy-preserving smart metering protocol for real-time tariff billing with dynamic policy adjustment. While our protocol preserves user privacy, it maintains the data utility required for accurate dynamic billing using real-time tariffs. Before describing our protocol, it is crucial to examine our system model, threat model, underlying assumptions, and design goals.
	
	\subsection {System Model}
\begin{center}
		\begin{figure}[htp]
			\includegraphics[width=3.5in]{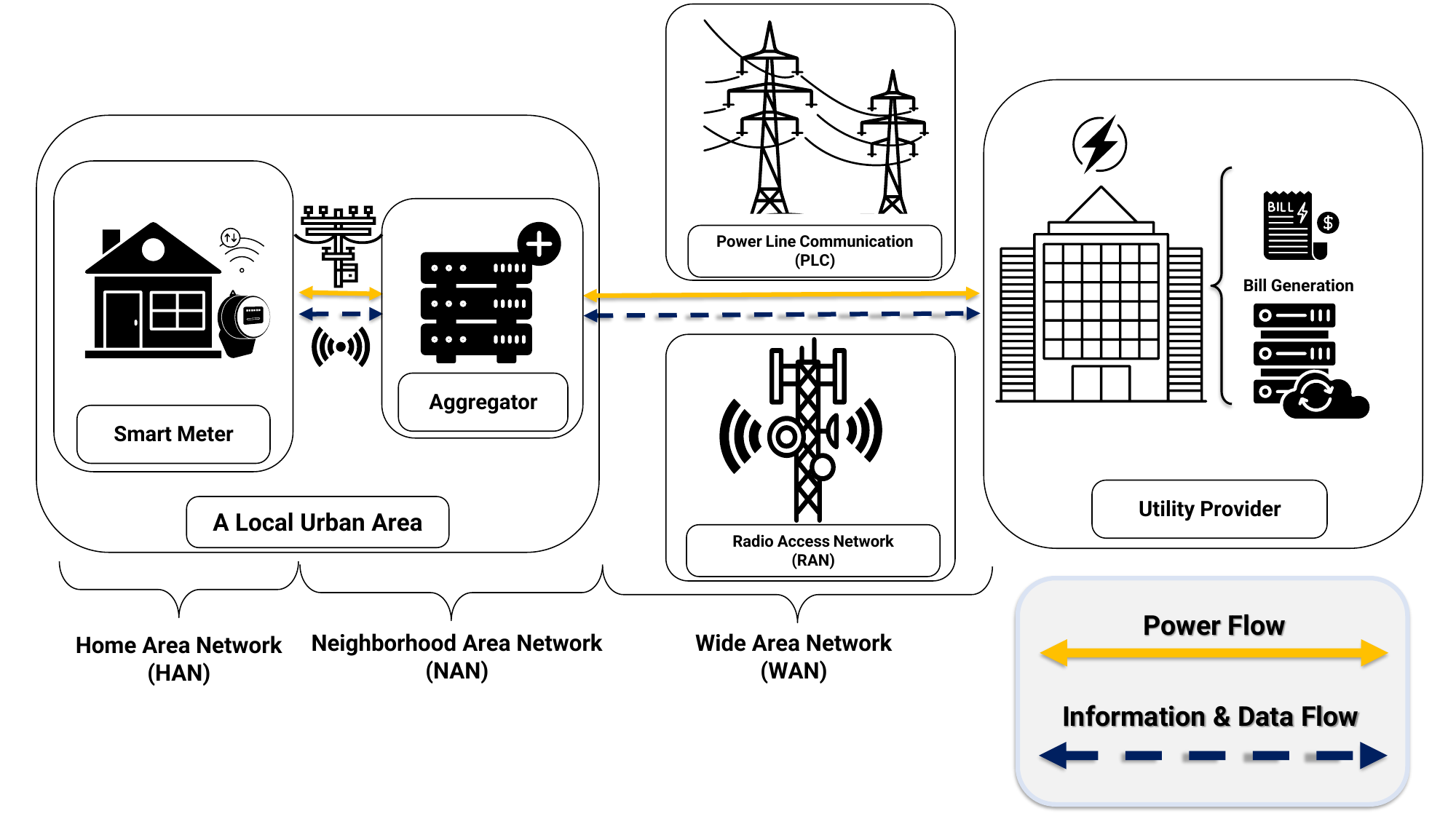}
			\caption{The smart grid's architecture integrates smart meters, aggregators, and the utility provider}
			\label{bill-system model}
		\end{figure}
	\end{center}

	Our system model defines a smart grid architecture comprising three main entities: (1) smart meters, (2) local aggregators, and (3) the utility provider. In the scheme's network model, each entity communicates with another entity using different communication technologies. For example, smart devices and home appliances connect to the smart meter via a home area network (HAN). Then each smart meter connects to its local aggregator through a neighborhood area network (NAN). Finally, local aggregators connect to the utility provider via a wide area network (WAN). In the following, we explain each entity and its role in the proposed scheme in more detail.
	\begin{enumerate}
		\item \textbf{Smart Meters}: Smart electricity meters measure fine-grained consumption readings at predefined intervals (e.g., every 15 minutes). In our protocol, smart meters perturb the measured consumption readings and report the resulting noisy consumption values to the utility provider via their local intermediary gateways, referred to as aggregators. Smart meters are computationally bounded and have low computational resources. It is important to notice that smart meters are capable of performing primitive cryptographic operations required by the protocol.
		
		\item \textbf{Local Aggregators}: Aggregators are considered as intermediary nodes that relay meter readings and protocol messages between smart meters and the utility provider. These local nodes have sufficient computing resources and also can be utilized as local gateways to handle outsourced computations. In our scheme, they only act as relay nodes.
		
		\item \textbf{Utility Provider}: The utility provider is responsible for providing various services to customers, including dynamic billing, grid and demand-supply management, and load forecasting. In our protocol, the utility provider specifies real-time tariffs, collects the reported noisy consumption values, and generates customers' bills. It may also initiate tariff policy adjustments when required.
\end{enumerate}
	The smart grid architecture and network model are shown in Fig. \ref{bill-system model}.
	\subsection {Threat Model and Assumptions}
	In the proposed scheme, we specify distinct levels of trust for each main entity based on the protocol's system model. Consequently, we can define our scheme's threat model as follows.
	\begin{itemize}
		\item \textbf{Smart Meters}: Smart meters are fully trusted and tamper-resistant. We assume that their trusted metrology and
		credential-storage components remain uncompromised during protocol execution. These intelligent power meters are equipped with
		tamper-protection mechanisms intended to detect or mitigate unauthorized physical manipulation. Successful physical compromise of the trusted metering components, including attacks that bypass these protections, is outside the scope of the proposed privacy-preserving billing protocol. 	
		\item \textbf{Aggregators}: Local aggregators are assumed to be honest but curious, which means they adhere to the protocol but may attempt to infer private information out of curiosity. Local aggregators are managed and configured by the utility provider; therefore, aggregators and the utility provider may collude by sharing the information they legitimately observe during protocol execution. Such colluding entities remain honest but curious and do not deviate from the protocol.
		\item \textbf{Utility Provider}: The utility provider is considered as a semi-trusted entity similar to local intermediary gateways discussed earlier. This entity follows the protocol but may analyze consumers' reported noisy consumption readings to infer fine-grained consumption information, usage patterns, or other private information for its own advantage.
	\end{itemize}
	Accordingly, the semi-trusted entities are assumed to know the public protocol parameters, tariff information, and the noisy consumption values legitimately available to them during protocol execution. They may also exploit available statistical background knowledge when analyzing these observations. However, they do not know the original fine-grained consumption values, the individual noise values generated by a smart meter, or the secret seed used by the PRNG. 
	
\begin{center}
		\begin{figure}[htp]
			\includegraphics[width=3.5in]{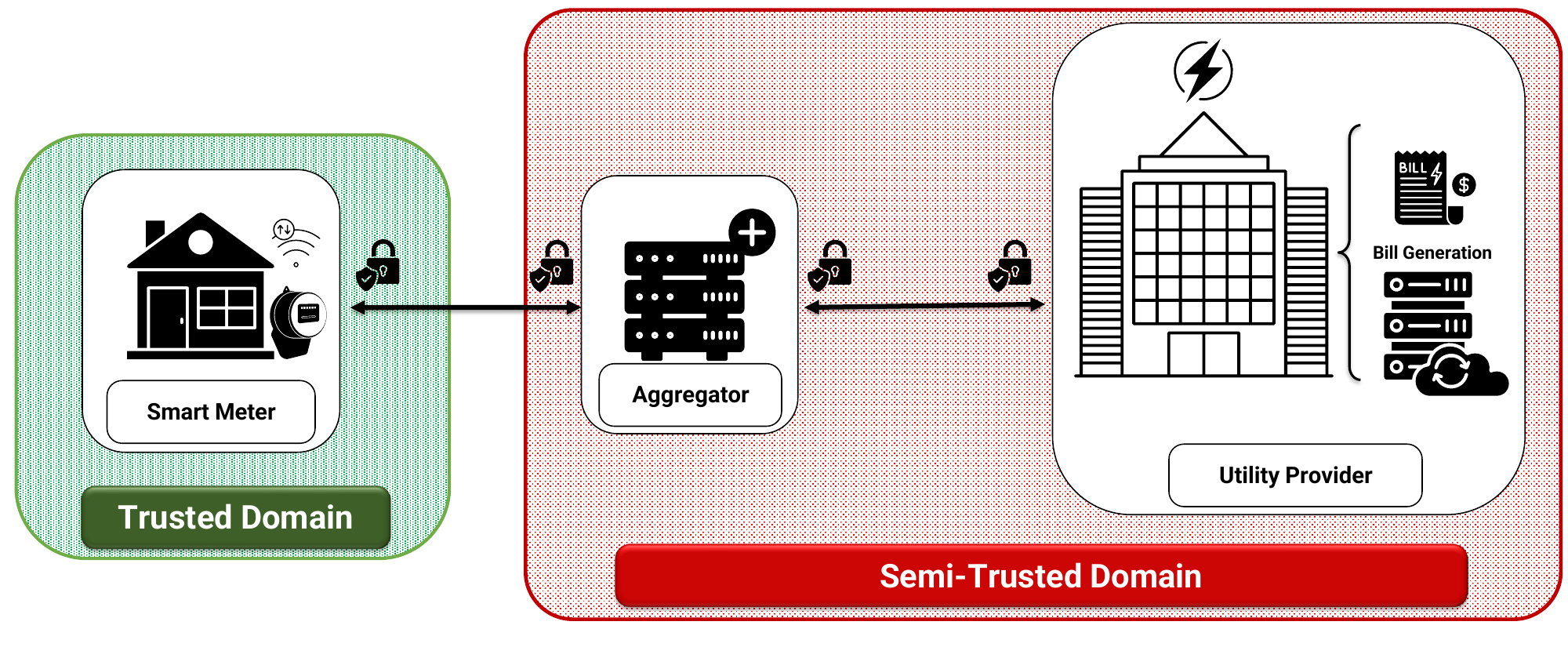}
			\caption{Our scheme's threat model}
			\label{bill-threat model}
		\end{figure}
	\end{center}

	Our scheme's threat model is shown in Fig. \ref{bill-threat model}. 

	Our protocol also establishes multiple assumptions relevant to the AMI infrastructure communication and its components, which are discussed as follows.
	\begin{enumerate} [a)]
		\item \textbf{Secure and Fortified Bidirectional Channels}: All bidirectional links between AMI components are assumed to be secured using TLS, thereby protecting transmitted protocol messages against passive eavesdropping and active modification or impersonation attacks. 
		\item \textbf{Privacy-Aware Pre-Authentication}: Prior to protocol initiation, AMI components use a privacy-aware authentication scheme to register themselves within the AMI network.
		\item \textbf{Robust Public-Key Dissemination}: Public keys are distributed securely among meters, aggregators, and the utility provider --- often through mechanisms like public key certificates --- ensuring their integrity throughout the process.
		\item \textbf{Authenticated Tariff Policy Updates}: Tariff information and any subsequent tariff policy updates are assumed to originate from the utility provider and are delivered to smart meters through authenticated protocol messages. Smart meters only process tariff policy updates that are validly authenticated within the AMI infrastructure.
	\end{enumerate}

	\subsection{Design Goals}	
	In our lightweight privacy-preserving smart metering billing protocol, we aim to achieve multiple goals as follows:
	\begin{enumerate} [I.]
		\item \textbf{Data Privacy}: The main goal is to preserve meter data privacy by incorporating lightweight privacy-preserving techniques such as data perturbation and data minimization. In the proposed scheme, a resource-friendly noise application mechanism is utilized to perturb consumption readings. As a consequence, only noisy consumption values are reported, thereby reducing the disclosure of attributable
		fine-grained consumption information to semi-trusted entities.
			
		\item \textbf{Billing Utility}: While our scheme perturbs attributable fine-grained metering data to preserve customers' privacy, it maintains the utility required for accurate billing through the tariff-weighted zero-sum noise property introduced by the noise application mechanism.
		
		\item \textbf{Lightweight Dynamic Accurate Billing}: Our scheme supports the real-time pricing (RTP) policy, which is a time-based dynamic pricing approach. This ensures that specified real-time tariffs are delivered to smart meters and informs customers about electricity costs at each interval based on grid status. Unlike privacy-preserving billing approaches that rely on computationally intensive cryptographic operations, our protocol enables accurate interval-based billing using lightweight data perturbation and basic arithmetic operations at smart meters, while preventing the direct disclosure of original fine-grained consumption readings.
		
		\item \textbf{Dynamic Tariff Policy Adjustment}: Upon any authorized tariff policy changes, the utility provider can recompute the users'
		bills according to the updated tariffs while reusing the previously reported noisy consumption values and requiring only limited additional
		processing and communication from smart meters.
	\end{enumerate}
	
	\section{Proposed Scheme}
	As we discussed earlier, attributable fine-grained consumption values are necessary to generate accurate dynamic bills for customers. Although fine-grained readings are required for bill generation and accounting, such time-series data can reveal consumers’ daily habits and behaviors throughout the day. To overcome these privacy challenges and provide a precise dynamic billing system based on real-time tariffs, we propose a lightweight, privacy-preserving smart metering protocol for real-time tariff billing with dynamic tariff policy adjustment. In this section, we present an overview of our protocol to offer a clear and concise introduction, followed by a formal and detailed explanation.

	\subsection{Protocol Overview}
	Our privacy-aware billing protocol is designed under two distinct scenarios. In the first scenario, tariff information is delivered at each interval to smart meters according to the adopted RTP policy, and no tariff policy adjustment is applied after the billing period. This is a typical scenario where the utility provider maintains the originally adopted tariff policy. However, in the second scenario, the utility provider may adjust the tariff policy after the billing period. For example, it may reduce electricity costs for certain intervals due to governmental or regulatory mandates. In this situation, the utility provider can recompute the bills and notify customers about the newly adopted tariff policy and newly generated bills.
	
	\noindent\textbf{Scenario I --- Standard Billing Without Tariff Policy Adjustment:}
	In this scenario, the utility provider maintains the originally adopted tariff policy throughout the billing period. This scenario consists of
	multiple steps, outlined below:
	
	\begin{enumerate}
		\item The utility provider initiates the protocol by specifying different tariffs for each urban or rural district within a city and transmits the defined real-time tariffs to local aggregators at each interval (e.g., every 15 minutes).
		\item Each local aggregator sends tariffs to the smart meters within its network to inform customers about the electricity costs corresponding to their local urban or rural area at that interval.
		\item Upon receiving the tariff information, each smart meter informs the customer about the electricity price at that interval via an in-home display. In the case of peak hours, this can help customers reduce their energy consumption and shift non-essential energy-consuming tasks to off-peak hours, thereby optimizing usage and saving money.
		\item Afterward, each smart meter generates a random value using a counter-based PRNG (i.e., ThreeFry) that follows a normal distribution with a mean of zero and a standard deviation of $\sigma$. These random values are retained for a specific duration in anticipation of potential tariff policy adjustments.
		\item It then perturbs the consumption reading of that interval by adding the original consumption data to the generated random value to generate a noisy consumption value. Subsequently, each meter sends the noisy consumption reading to the utility provider.
		\item Simultaneously, each meter retains the generated random value for possible tariff policy adjustments, multiplies it by the tariff received
		for that period, and stores the resulting tariff-weighted value.
		\item The process continues through successive intervals until the one immediately preceding the bill generation period. In the last interval, the smart meter performs a summation operation over the stored multiplied results. Then, each meter computes the additive inverse of the resulting sum and divides it by the last received tariff to generate a correction noise value for the last consumption data.
		\item It then perturbs the last power usage data and sends the final noisy reading to the utility provider.
		\item The utility provider only collects the noisy consumption values over a billing period (e.g., a week, two weeks, or a month). As we mentioned earlier, our noise mechanism adheres to a zero-sum property that leads to accurate dynamic bills for each customer. As a result, the utility provider can generate bills using noisy consumption values.
		\item The utility provider multiplies each noisy consumption value by the corresponding tariff and generates a vector of noisy partial bills.
		\item Subsequently, it performs a summation over the noisy partial bills and computes the final bills for each customer for that specified billing period, leveraging the zero-sum property of the applied noise mechanism to ensure accuracy.
	\end{enumerate}
	
	\noindent\textbf{Scenario II --- Billing With Tariff Policy Adjustment:} In the second scenario, due to an authorized tariff policy adjustment,
	the utility provider must recalculate the bills using a new tariff vector. Depending on how the new tariff policy differs from the originally adopted tariff vector, we distinguish between two types of tariff policy adjustments: proportional and non-proportional adjustments.
		
 	\textbf{Type I --- Proportional Tariff Adjustment}: In this case, all tariffs are modified by a common positive scaling factor $\alpha$, such that $TRF_{new}=\alpha TRF$. Since proportional scaling preserves the tariff-weighted zero-sum property of the applied noise values, the utility	provider can recompute the new bill using the previously reported noisy consumption values without requiring the smart meter to generate or report a new final noisy consumption value.
		
	\textbf{Type II --- Non-Proportional Tariff Adjustment}: In this case, some tariff elements are modified non-uniformly such that no common scaling factor $\alpha$ satisfies $TRF_{new}=\alpha TRF$. Consequently, the previously generated final noise value may no longer satisfy the
	tariff-weighted zero-sum property under the new tariff vector. Therefore, a new correction noise value must be generated for the final consumption reading. This adjustment is composed of several steps, discussed below.
\begin{enumerate}
	\item At the end of the billing period, when the last noisy consumption value is reported, the utility provider may be required to change the
	adopted tariff policy (e.g., due to regulatory mandates).
	
	\item In this situation, the utility provider forwards a new tariff vector, which consists of tariff information for different intervals,
	to smart meters through local aggregators for urban or rural areas.
	
	\item Upon receiving an authenticated new tariff vector, each smart meter validates the update according to the protocol requirements. Since repeated non-proportional tariff adjustments may introduce additional billing relations over the same fine-grained consumption data, they can increase the information available to the utility provider. Therefore, as a privacy safeguard, the number of Type II adjustments for each billing period is limited to $K_{\max}$. The smart meter processes the adjustment only if $K<K_{\max}$; otherwise, no further Type II adjustment is processed for that billing period.
	
	\item If the adjustment is accepted, each smart meter multiplies each stored random value by the corresponding new tariff, then performs a
	summation over the resulting values. It then computes the additive inverse of the resulting sum, which is divided by the last new tariff,
	to generate a new correction noise value. 
	
	\item It perturbs only the final consumption data using the calculated correction noise value and transmits the final noisy consumption value
	to the utility provider.
	
	\item The utility provider applies the new tariffs to the noisy consumption values and aggregates the noisy partial bills. Due to the
	restored zero-sum property of the data perturbation mechanism, the noise values cancel out, and a new final bill is generated for each customer.
\end{enumerate}

	According to our protocol, we enable utility providers to dynamically generate consumers' accurate bills using the RTP policy in a secure and
	privacy-preserving way. Now we delve deeper into the normal billing scenario and the two tariff-policy adjustment types and explain the
	proposed billing scheme in detail. 

	\subsection{Protocol Details}
	We now explain each privacy-preserving billing scenario formally. The symbols utilized in the formal specification of the protocol are presented in Table \ref{table:element_description_bill}. 
	\begin{table}[htp]
		\centering
		\caption{Notation Used in the Proposed Billing Protocol}
		\label{table:element_description_bill}
		\footnotesize
		\renewcommand{\arraystretch}{1.25}
		\setlength{\tabcolsep}{3pt}
		
		\begin{tabularx}{\columnwidth}{
				>{\centering\arraybackslash}p{0.10\columnwidth}
				>{\raggedright\arraybackslash}p{0.47\columnwidth}
				>{\raggedright\arraybackslash}X
			}
			\hline
			\textbf{Symbol} &
			\textbf{Description} &
			\textbf{Formal Representation} \\
			\hline
			
			\(M\) &
			Number of registered and active smart meters in the network &
			\(M \in \mathbb{N}\) \\
			\hline
			
			\(L\) &
			Number of intervals during a billing period &
			\(L \in \mathbb{N}\) \\
			\hline
			
			\(\lambda\) &
			Number of days in a billing period &
			\(\lambda \in \mathbb{N}\) \\
			\hline
			
			\(t_i\) &
			The \(i\)-th time interval &
			\(T=\langle t_i\rangle_{i=1}^{L}\) \\
			\hline
			
			\(c_{t_i}\) &
			The consumption data at the \(i\)-th time interval &
			\(C=\langle c_{t_i}\rangle_{i=1}^{L}\) \\
			\hline
			
			\(trf_{t_i}\) &
			The tariff information/electricity cost at the \(i\)-th time interval &
			\(TRF=\langle trf_{t_i}\rangle_{i=1}^{L}\) \\
			\hline
			
			\(s_{t_i}\) &
			The noise value applied at the \(i\)-th time interval; the first
			\(L-1\) values are generated using a secure PRNG, while \(s_{t_L}\)
			is computed as the correction noise value &
			\(S=\langle s_{t_i}\rangle_{i=1}^{L}\) \\
			\hline
			
			\(nc_{t_i}\) &
			The noisy consumption data at the \(i\)-th time interval &
			\(NC=\langle nc_{t_i}\rangle_{i=1}^{L}\) \\
			\hline
			
			\(nb_{t_i}\) &
			The noisy partial bill at the \(i\)-th time interval \(t_i\) &
			\(NB=\langle nb_{t_i}\rangle_{i=1}^{L}\) \\
			\hline
			
			\(FB_{id_j}\) &
			The issued final bill for a smart meter with an ID of \(id_j\) &
			\(FB=\{FB_{id_j}\}_{j=1}^{M}\) \\
			\hline
			
			\(\alpha\) &
			Common positive scaling factor used in a proportional tariff
			policy adjustment &
			\(\alpha\in\mathbb{R}_{>0}\),
			\(trf'_{t_i}=\alpha trf_{t_i}\) \\
			\hline
			
			\(K\) &
			Number of accepted non-proportional tariff policy adjustments
			for the current billing period &
			\(K\in\mathbb{N}_0\) \\
			\hline
			
			\(K_{\max}\) &
			Maximum permitted number of non-proportional tariff policy
			adjustments during a billing period &
			\(K_{\max}\in\mathbb{N}\) \\
			\hline
		\end{tabularx}
	\end{table}

	At the beginning of each billing period, the smart meter initializes the counter associated with that billing period for accepted non-proportional tariff policy adjustments as $K\leftarrow0$.
	
	\noindent\textbf{Scenario I --- Standard Billing Without Tariff Policy Adjustment:} In our protocol, smart meters report perturbed values every 15 minutes to the utility provider. Based on the length of reporting intervals $|t_i|$ (i.e. 15-minute intervals) and the number of days in a billing period $\lambda$, the number of reports/intervals during a billing period is computed as follows:
	\begin{IEEEeqnarray}{rCl}
		L = \left( \left \lceil \dfrac{24 \cdot 60}{|t_i|} \right \rceil  \right) \cdot \lambda  
		\label{eq:bill-protocol-details-number-intervals}
	\end{IEEEeqnarray}
	For instance, a smart meter configured to transmit data every 15 minutes is expected to send a total of 96 reports per day.
	
	The utility provider specifies tariffs for the electricity unit (e.g., \$0.05/kWh) at each interval to offer real-time tariffs based on grid status, each area’s history, and regional characteristics (e.g., rural, urban, and industrial regions are charged based on different tariff policies). In fact, these provisioned real-time tariffs are broadcast to smart meters in each area through local aggregators. 
	 
	Upon receiving the tariff, the smart meter generates a random value using a secure PRNG with a secure initial parameter $seed$. The generated
	random value follows a Gaussian distribution with a zero mean and a standard deviation of $\sigma$.
	\begin{IEEEeqnarray}{rCl}
		s_{t_i} &\leftarrow& RandGen(seed,\sigma), \notag\\
		s_{t_i} &\overset{\mathrm{i.i.d.}}{\sim}&
		\mathcal{N}(0,\sigma^2),
		\quad 1 \leq i \leq L-1
		\label{eq:bill-protocol-randomGen}
	\end{IEEEeqnarray}
	Then the smart meter perturbs the recorded consumption reading 	$c_{t_i}$ as follows:
	\begin{IEEEeqnarray}{rCl}
		nc_{t_i} \leftarrow c_{t_i} + s_{t_i},
		\quad 1 \leq i \leq L-1
		\label{eq:bill-protocol-perturb-cons-data}
	\end{IEEEeqnarray}
	
	Afterward, it sends noisy consumption values at each interval through a secure channel to the utility provider via its local intermediary gateway.
	Simultaneously, the smart meter retains the generated noise value $s_{t_i}$ for possible tariff policy adjustments and stores its tariff-weighted value required for the final correction-noise computation.
	\begin{IEEEeqnarray}{rCl}
		Store(s_{t_i},\,s_{t_i}\cdot trf_{t_i}),
		\quad 1 \leq i \leq L-1
		\label{eq:bill-protocol-tariff-rand-mult}
	\end{IEEEeqnarray}
	This process continues through time interval $t_{L-1}$.
 	
	As we previously discussed, our data perturbation mechanism takes advantage of a tariff-weighted zero-sum property. By using this noise
	cancellation feature, the utility provider can generate accurate dynamic bills for each customer using the noisy consumption readings reported
	during the billing period. This tariff-weighted zero-sum property is defined as follows:
  	\begin{IEEEeqnarray}{rCl}
 		\sum_{i=1}^{L} (s_{t_i} \cdot trf_{t_i}) = 0
 		\label{eq:bill-protocol-zero-sum-property-noise-mechanism}
 	\end{IEEEeqnarray}
 	
	To achieve the tariff-weighted zero-sum property defined in Relation~\ref{eq:bill-protocol-zero-sum-property-noise-mechanism}, the smart meter computes a correction noise value using the stored values and the tariff of the final interval. This computation requires $trf_{t_L} \neq 0$. The correction noise value applied to the final consumption reading is computed as follows:
  	\begin{IEEEeqnarray}{rCl}
 		s_{t_L} = -\dfrac{\sum_{i=1}^{L-1} (s_{t_i} \cdot trf_{t_i})}{trf_{t_L}} 
 		\label{eq:bill-protocol-last-noise-generation}
 	\end{IEEEeqnarray}
	Now, the smart meter applies the computed correction noise value to the last power usage data and sends the resulting noisy consumption value to the utility provider.
	\begin{IEEEeqnarray}{rCl}
		nc_{t_L} \leftarrow c_{t_L} + s_{t_L}
		\label{eq:bill-protocol-perturbed-last-reading}
	\end{IEEEeqnarray}
	The smart meter retains $c_{t_L}$ together with the required noise values for the permitted tariff policy adjustment period.
 	
	Upon receiving the last noisy consumption data, the utility provider starts to compute the user's bill by multiplying each perturbed consumption value by its corresponding tariff as follows:
	\begin{IEEEeqnarray}{rCl}
		nb_{t_i} &\leftarrow& nc_{t_i} \cdot trf_{t_i},
		\quad 1 \leq i \leq L \notag \\
		NB &=& \langle nb_{t_1}, nb_{t_2}, \ldots, nb_{t_L} \rangle
		\label{eq:bill-multiply-noisy-data-to-tariff}
	\end{IEEEeqnarray}
	Subsequently, the utility provider performs a summation operation over the noisy partial bills to generate the final electricity bill for the customer.
  	\begin{IEEEeqnarray}{rCl}
		FB_{id_j} = \sum_{i=1}^{L} nb_{t_i}
		\label{eq:bill-electricty-bill-generation}
	\end{IEEEeqnarray} 

	\noindent\textbf{Proposition 1 (Correctness of Scenario I):} Given $trf_{t_L}\neq0$ and the correction noise value $s_{t_L}$ computed according to Relation~\ref{eq:bill-protocol-last-noise-generation}, the final bill computed from the noisy consumption values is identical to the bill obtained from the original consumption values.
	
\begin{center}
		\begin{figure}[htp]
			\includegraphics[width=3.5in]{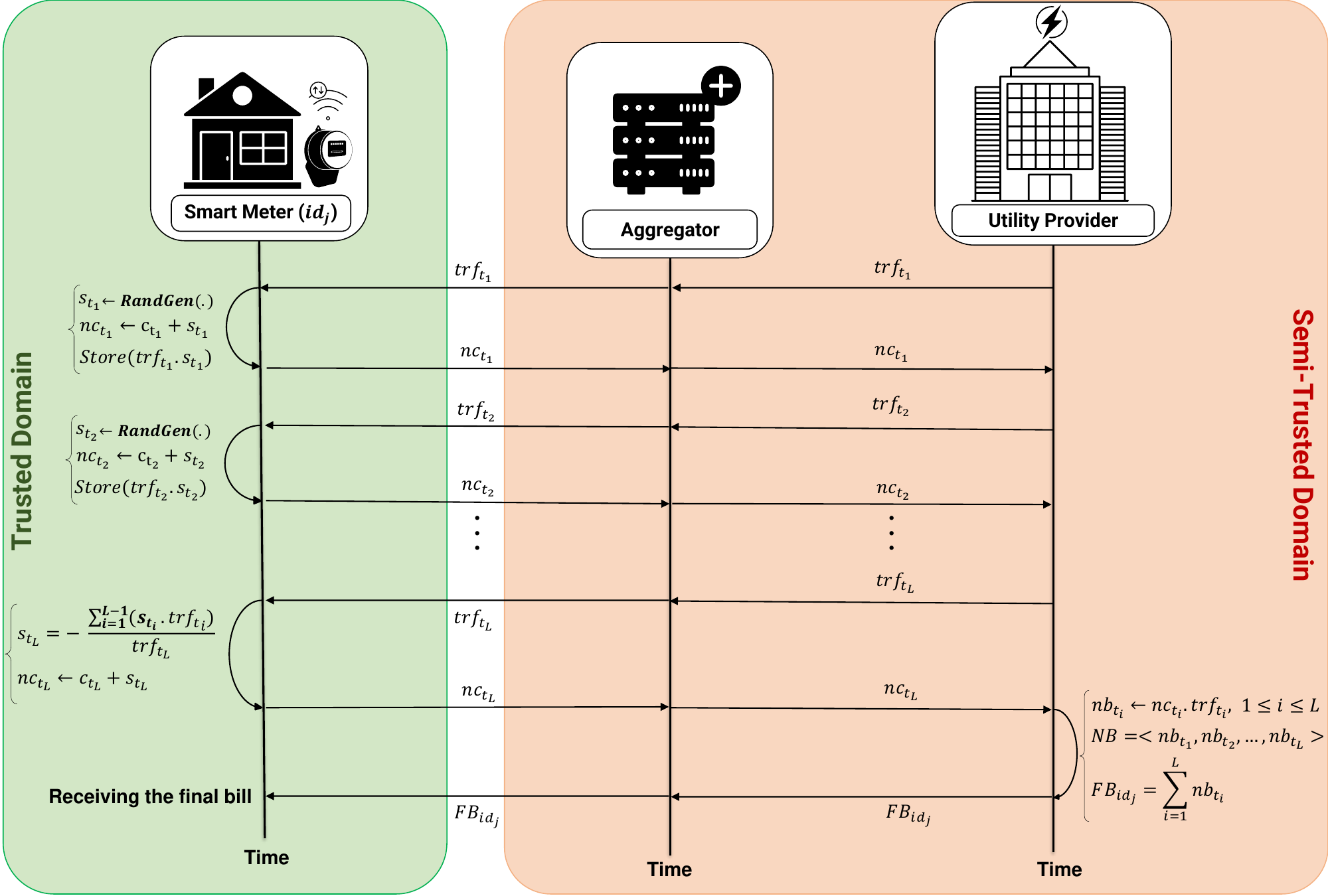}
			\caption{Protocol flow for Scenario I: standard billing without tariff policy adjustment}
			\label{bill-protocol-details-I}
		\end{figure}
	\end{center}

	\noindent\textit{Proof.} The correctness of Scenario I is demonstrated by showing that the tariff-weighted noise terms cancel exactly in the final billing computation:
  	\begin{IEEEeqnarray}{rCl}
		FB_{id_j} &=& \sum_{i=1}^{L} nb_{t_i} \notag \\
		          &=& \sum_{i=1}^{L} (nc_{t_i} \cdot trf_{t_i}) \notag \\
		          &=& \sum_{i=1}^{L} ((c_{t_i} + s_{t_i}) \cdot trf_{t_i}) \notag \\
      		      &=& \sum_{i=1}^{L} \left( (c_{t_i} \cdot trf_{t_i}) + (s_{t_i}\cdot trf_{t_i}) \right)\notag \\
      		      &=& \sum_{i=1}^{L} (c_{t_i} \cdot trf_{t_i}) + \sum_{i=1}^{L} (s_{t_i}\cdot trf_{t_i})\notag \\
      		      &=& \sum_{i=1}^{L} (c_{t_i} \cdot trf_{t_i}) + \sum_{i=1}^{L-1} (s_{t_i}\cdot trf_{t_i}) + (s_{t_L}\cdot trf_{t_L}) \notag \\
      		      &=& \sum_{i=1}^{L} (c_{t_i} \cdot trf_{t_i}) + \sum_{i=1}^{L-1} (s_{t_i}\cdot trf_{t_i}) - \sum_{i=1}^{L-1} (s_{t_i}\cdot trf_{t_i}) \notag \\
      		      &=& \sum_{i=1}^{L} (c_{t_i} \cdot trf_{t_i}) 
		\label{eq:bill-noise cancellation-process-I}
	\end{IEEEeqnarray} 
	Therefore, the tariff-weighted noise contribution is exactly canceled, and the final bill computed from the noisy consumption values is identical
	to the bill obtained from the original consumption values. This proves the billing correctness of Scenario I. The protocol flow for Scenario I is depicted in
	Fig.~\ref{bill-protocol-details-I}.

	\noindent\textbf{Scenario II --- Billing With Tariff Policy Adjustment:} In the second scenario, the utility provider may be required to recompute the meter's final bill due to governmental or regulatory mandates after the billing period. In such cases, the utility provider provisions a new tariff vector
	
	\begin{IEEEeqnarray}{rCl}
		TRF_{new}
		&=&
		\langle
		trf_{t_1}^{'}, trf_{t_2}^{'}, \ldots, trf_{t_L}^{'}
		\rangle
		\label{eq:bill-new-tariff-vector}
	\end{IEEEeqnarray}
	and broadcasts it to the smart meters within the corresponding area. Depending on the relation between $TRF_{new}$ and the originally adopted
	tariff vector $TRF$, the tariff policy adjustment is classified as either a proportional (Type I) or a non-proportional (Type II) adjustment.
	
	\noindent\textbf{Type I --- Proportional Tariff Adjustment:} A proportional tariff adjustment occurs when all elements of the new tariff
	vector are modified by the same positive scaling factor $\alpha$, such that
	\begin{IEEEeqnarray}{rCl}
		trf_{t_i}^{'} &=& \alpha \cdot trf_{t_i},
		\quad 1 \leq i \leq L,
		\qquad \alpha \in \mathbb{R}_{>0}
		\label{eq:bill-proportional-tariff-adjustment}
	\end{IEEEeqnarray}
	In this case, the previously generated noise values preserve the tariff-weighted zero-sum property under the new tariff vector. Specifically,
	\begin{IEEEeqnarray}{rCl}
		\sum_{i=1}^{L}
		(s_{t_i} \cdot trf_{t_i}^{'})
		&=&
		\sum_{i=1}^{L}
		(s_{t_i} \cdot \alpha trf_{t_i}) \notag \\
		&=&
		\alpha
		\sum_{i=1}^{L}
		(s_{t_i} \cdot trf_{t_i}) \notag \\
		&=& 0.
		\label{eq:bill-proportional-zero-sum}
	\end{IEEEeqnarray}
	Therefore, the correction noise value also remains unchanged under a proportional tariff adjustment:
	\begin{IEEEeqnarray}{rCl}
		s_{t_L}^{'}
		&=&
		-\dfrac{
			\sum_{i=1}^{L-1}
			(s_{t_i} \cdot trf_{t_i}^{'})
		}{
			trf_{t_L}^{'}
		} \notag \\
		&=&
		-\dfrac{
			\sum_{i=1}^{L-1}
			(s_{t_i} \cdot \alpha trf_{t_i})
		}{
			\alpha trf_{t_L}
		} \notag \\
		&=& s_{t_L}.
		\label{eq:bill-proportional-correction-noise}
	\end{IEEEeqnarray}
	
	\noindent\textbf{Corollary 1 (Correctness of Type I Tariff Adjustment):} Consequently, no new correction noise value or additional noisy consumption
	report is required from the smart meter. The utility provider can directly recompute the adjusted bill using the previously reported noisy consumption
	values and the new tariff vector.

	\noindent\textit{Proof.}
	The correctness of the Type I tariff adjustment follows directly from the preserved tariff-weighted zero-sum property established in  	Relation~\ref{eq:bill-proportional-zero-sum}. Therefore, the adjusted final bill is computed as
	\begin{IEEEeqnarray}{rCl}
	FB_{id_j}^{new}
	&=&
	\sum_{i=1}^{L}
	(nc_{t_i}\cdot trf_{t_i}^{'}) \notag \\
	&=&
	\sum_{i=1}^{L}
	(c_{t_i}\cdot trf_{t_i}^{'})
	\label{eq:bill-proportional-adjusted-bill}
	\end{IEEEeqnarray}
	
	Therefore, the adjusted bill computed from the previously reported noisy consumption values is identical to the bill obtained from the original
	consumption values under the new tariff vector.
	
	Since Type I adjustments do not require the generation of a new correction noise value or a new noisy report, they do not increment the
	non-proportional adjustment counter $K$.
	
	\noindent\textbf{Type II --- Non-Proportional Tariff Adjustment:} A non-proportional tariff adjustment occurs when no common scaling factor
	$\alpha$ satisfies $TRF_{new}=\alpha TRF$. In this case, the previously computed correction noise value may no longer satisfy the tariff-weighted
	zero-sum property under the new tariff vector. For each accepted Type II adjustment, the prime notation denotes the currently adjusted tariff vector
	and its corresponding recomputed correction noise and final noisy consumption value; subsequent accepted adjustments follow the same procedure.
	
	Upon receiving an authenticated non-proportional tariff policy adjustment, the smart meter validates the received tariff vector and verifies that
	$K<K_{\max}$. Moreover, since the last updated tariff is used as the denominator in the correction-noise computation, the adjustment requires
	$trf_{t_L}^{'}\neq0$. If these conditions are not satisfied, the smart meter rejects the adjustment request.
	
	If the adjustment is accepted, the smart meter loads the random values generated during the billing period, multiplies each by the corresponding new tariff $trf_{t_i}^{'}$, and performs a summation over the resulting values. Then, the additive inverse of the resulting value divided by the last new tariff $trf_{t_L}^{'}$ is computed to generate a new correction noise value:
	\begin{IEEEeqnarray}{rCl}
		s_{t_L}^{'}
		&=&
		-\dfrac{
			\sum_{i=1}^{L-1}
			(s_{t_i} \cdot trf_{t_i}^{'})
		}{
			trf_{t_L}^{'}
		}
		\label{eq:bill-noise-generation-after-policy-change}
	\end{IEEEeqnarray}
	
	The smart meter then perturbs the last consumption data using the newly computed correction noise value and forwards the resulting noisy consumption value to the utility provider:
	\begin{IEEEeqnarray}{rCl}
		nc_{t_L}^{'}
		&=&
		c_{t_L}+s_{t_L}^{'}
		\label{eq:bill-apply-noise-to-last-reading-after-policy-change}
	\end{IEEEeqnarray}
	After successfully processing the non-proportional tariff adjustment, the smart meter increments the corresponding adjustment counter as
	$K\leftarrow K+1$.
	
	Since the new final perturbed reading enables the utility provider to recompute the customer's final bill, the utility provider replaces $nc_{t_L}$ with $nc_{t_L}^{'}$ and retains and reuses the previously reported $L-1$ noisy consumption readings. The new vector of noisy partial bills and the corresponding final bill are then recomputed as follows:	
	\begin{IEEEeqnarray}{rCl}
		nb_{t_i}^{'}
		&=&
		nc_{t_i}\cdot trf_{t_i}^{'},
		\quad 1 \leq i \leq L-1
		\notag \\
		nb_{t_L}^{'}
		&=&
		nc_{t_L}^{'}\cdot trf_{t_L}^{'}
		\notag \\
		NB^{'}
		&=&
		\langle
		nb_{t_1}^{'}, nb_{t_2}^{'}, \ldots, nb_{t_L}^{'}
		\rangle
		\notag \\
		FB_{id_j}^{new}
		&=&
		\sum_{i=1}^{L} nb_{t_i}^{'}
		\label{eq:bill-final-bill-generation-after-policy-change}
	\end{IEEEeqnarray}
	
	\noindent\textbf{Proposition 2 (Correctness of Type II Tariff Adjustment):}
	For an accepted non-proportional tariff policy adjustment with $trf_{t_L}^{'}\neq0$ and the correction noise value $s_{t_L}^{'}$
	computed according to Relation~\ref{eq:bill-noise-generation-after-policy-change}, the adjusted final bill computed from the noisy consumption values is identical to the bill obtained from the original consumption values under the new tariff vector.
	
	\noindent\textit{Proof.}
	Starting from Relation~\ref{eq:bill-noise-generation-after-policy-change} and multiplying both sides by $trf_{t_L}^{'}$ gives
	\begin{IEEEeqnarray}{rCl}
		s_{t_L}^{'}\cdot trf_{t_L}^{'}
		&=&
		-\sum_{i=1}^{L-1}
		(s_{t_i}\cdot trf_{t_i}^{'})
		\notag
	\end{IEEEeqnarray}
	and therefore,
	\begin{IEEEeqnarray}{rCl}
		\sum_{i=1}^{L-1}
		(s_{t_i}\cdot trf_{t_i}^{'})
		+
		s_{t_L}^{'}\cdot trf_{t_L}^{'}
		&=& 0.
		\label{eq:bill-adjusted-zero-sum}
	\end{IEEEeqnarray}
	
	Thus, the newly computed correction noise value restores the tariff-weighted zero-sum property under the adjusted tariff vector. Accordingly, the utility provider computes the adjusted final bill using the previously reported $L-1$ noisy consumption values and the newly reported final noisy consumption value as follows:
	\begin{IEEEeqnarray}{rCl}
		FB_{id_j}^{new}
		&=&
		\sum_{i=1}^{L-1}
		(nc_{t_i}\cdot trf_{t_i}^{'})
		+
		nc_{t_L}^{'}\cdot trf_{t_L}^{'}
		\notag \\
		&=&
		\sum_{i=1}^{L-1}
		((c_{t_i}+s_{t_i})\cdot trf_{t_i}^{'})
		+
		(c_{t_L}+s_{t_L}^{'})\cdot trf_{t_L}^{'}
		\notag \\
		&=&
		\sum_{i=1}^{L}
		(c_{t_i}\cdot trf_{t_i}^{'})
		+
		\sum_{i=1}^{L-1}
		(s_{t_i}\cdot trf_{t_i}^{'})
		+
		s_{t_L}^{'}\cdot trf_{t_L}^{'}
		\notag \\
		&=&
		\sum_{i=1}^{L}
		(c_{t_i}\cdot trf_{t_i}^{'})
		\label{eq:bill-adjusted-correctness}
	\end{IEEEeqnarray}
	
\begin{center}
		\begin{figure}[htp]
			\includegraphics[width=3.5in]{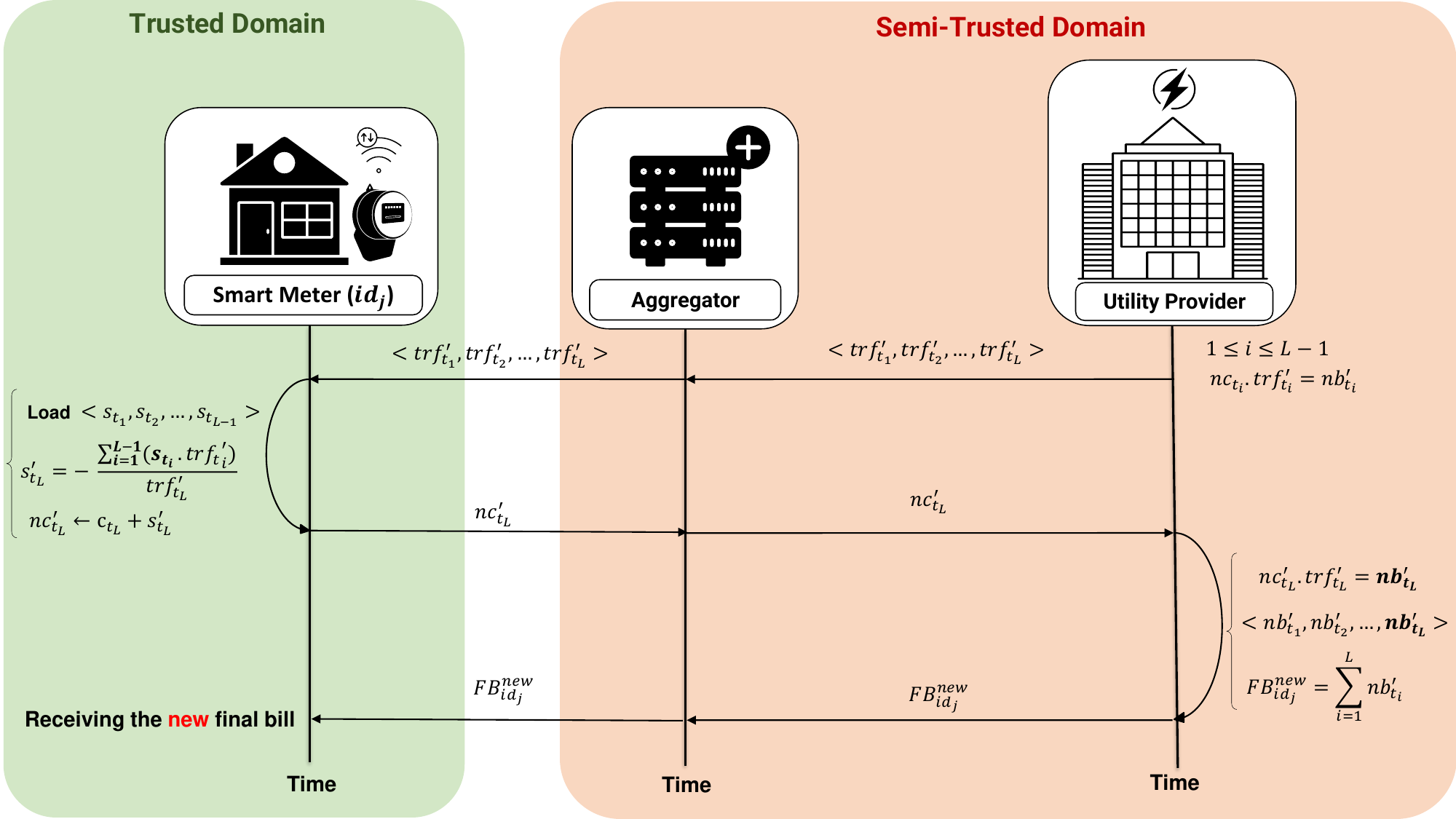}
			\caption{Protocol flow for a Type II non-proportional tariff policy adjustment}
			\label{bill-protocol-details-II}
		\end{figure}
	\end{center}

	Therefore, the restored tariff-weighted zero-sum property ensures that
	the adjusted final bill computed from the noisy consumption values is
	identical to the bill obtained from the original consumption values under
	the new tariff vector. Thus, the proposed Type II tariff adjustment
	preserves exact billing correctness without requiring the disclosure of
	the original fine-grained consumption readings. The protocol flow for a Type II non-proportional tariff policy adjustment
	is depicted in Fig.~\ref{bill-protocol-details-II}.
	
	\section{Evaluation}
	We evaluate our scheme from two distinct aspects: performance and privacy. 	To be more precise, we evaluate the scheme's performance using three different metrics, including computational, communication, and memory overhead. The performance evaluation combines analytical characterization and experimental assessment to examine the resource requirements of the proposed protocol.
	
	For the assessment of the scheme's privacy, we analyze the statistical properties of the applied noise and evaluate the extent to which the reported noisy consumption values protect the original fine-grained consumption profile against inference. We also analyze the information exposed by repeated non-proportional tariff policy adjustments. In addition, we retain the Jensen--Shannon (JS) divergence metric to measure the distributional difference between the original and perturbed consumption data.
	
	\subsection{Performance Evaluation}
	As we previously discussed, our performance evaluation consists of evaluating the computational, memory, and communication overhead. We combine analytical evaluation with experimental measurements to assess the computational and resource requirements of the proposed billing protocol.
	
	In our study, we implemented our scheme on two distinct computing platforms. To emulate the smart meter, we utilized the Quick Emulator (QEMU) to recreate the hardware environment of an Orange Pi One PC, featuring a 1.2GHz 32-bit ARM Cortex-A7 processor and 1GB of RAM. An Armbian Ubuntu-based operating system, which is specifically designed for Internet of Things (IoT) applications, runs on this virtualized setup. To represent the aggregator and the utility provider, the second platform was equipped with a 1.8GHz Intel Core i7 processor (8 cores) and 32GB of RAM. To investigate the protocol’s performance under constrained memory conditions, we intentionally restricted the RAM on this system to 1GB. It is also important to note that this machine ran on a Debian 12 operating system.
	
	The protocol was developed in Python. We employed the \texttt{secrets} module to generate the initial secret seeds and the \texttt{randomgen} library to generate pseudorandom values using the ThreeFry counter-based PRNG. NumPy was used to perform the required numerical computations.

	We based our analysis on a dataset from Slovakia’s AMI \cite{cenky2023dataset}, which includes energy consumption records from 1,000 anonymized residential smart meters. The dataset captures both active and reactive energy every 15 minutes over the course of a year. The dataset is complete, with no missing values or issues related to sparsity.
	
	\subsubsection{Computational Overhead} To start our analysis on the scheme’s computational overhead, we present a notation table (see Table \ref{table:bill-evaluation-computational-overhead-notation}), which lists formal symbols used in the analytical evaluation.
	\begin{table}[htp]
			\centering
			\caption{Notation Table for Computational Overhead}
			\normalsize
			\renewcommand{\arraystretch}{1.5} 
			\setlength{\tabcolsep}{6pt} 
			\resizebox{1.00\columnwidth}{!}{ 
					\begin{tabular}{|p{3.5cm}|p{9cm}|}
							\hline
							\textbf{Symbol} & \textbf{Description} \\
							\hline
							$\mathbf{t}_{PRNG}$ & Time of generating a random value using a secure PRNG  \\
							\hline
							$\mathbf{t}_{arthm}$ & Time of performing a basic arithmetic operation \\
							\hline
						\end{tabular}
				}
			\label{table:bill-evaluation-computational-overhead-notation}
		\end{table}
	
	According to the proposed scheme, each meter only performs lightweight operations, including random value generation, data perturbation, tariff-weighted arithmetic operations, and generating the final correction noise value that results in noise cancellation. $\mathbf{t}_{sm}$ represents the smart meter's execution time, which is computed as follows, where $L$ is the number of intervals during the billing period.
	\begin{IEEEeqnarray}{rCl}
		\mathbf{t}_{sm}
		&=&
		(L-1)\mathbf{t}_{PRNG}
		+
		(3L-1)\mathbf{t}_{arthm}
		\label{eq:bill-protocol-evaluation-comp-overhead-sm}
	\end{IEEEeqnarray}
	As we discussed earlier, aggregators act as intermediary nodes that only relay messages between smart meters and the utility provider. The utility provider performs two core operations, including computing noisy partial bills and adding them together to generate the final bill for the customer. The analytical computational overhead for the utility provider is as follows.
	\begin{IEEEeqnarray}{rCl}
		\mathbf{t}_{up}
		&=&
		(2L-1)\mathbf{t}_{arthm}
		\label{eq:bill-protocol-evaluation-comp-overhead-up}
	\end{IEEEeqnarray}
	Finally, the total protocol computational overhead is computed as follows:
	\begin{IEEEeqnarray}{rCl}
		\mathbf{t}_{bill}
		&=&
		\mathbf{t}_{sm}+\mathbf{t}_{up}
		\label{eq:bill-protocol-evaluation-comp-overhead}
	\end{IEEEeqnarray}
	The additional computational overhead introduced by a tariff policy adjustment depends on its type. In a Type I proportional adjustment, the smart meter does not generate a new correction noise value or a new noisy consumption report. Therefore, no additional billing computation is required at the smart meter. The utility provider recomputes the adjusted 	bill using the previously reported noisy consumption values, requiring $(2L-1)\mathbf{t}_{arthm}$.
	
	For each accepted Type II non-proportional adjustment, the smart meter 	multiplies the stored $L-1$ noise values by their corresponding new tariffs, performs the required summation, computes the new correction noise value, and perturbs the final consumption reading. Therefore, the additional computational overhead at the smart meter is
	\begin{IEEEeqnarray}{rCl}
		\mathbf{t}_{sm}^{II}
		&=&
		(2L)\mathbf{t}_{arthm}.
		\label{eq:bill-type-II-comp-overhead-sm}
	\end{IEEEeqnarray}
	
	The utility provider then recomputes the adjusted final bill with an additional computational overhead of $(2L-1)\mathbf{t}_{arthm}$. 
	
	Furthermore, we performed an experiment to assess the performance and efficiency of the proposed solution under the standard billing scenario. Table \ref{table:billing-service-evaluation} indicates the protocol execution time for each processing entity, including the smart meter and the utility provider. As indicated, the execution time for one month, the first three months, the first six months, the first nine months, and the entire year is computed.
	\begin{table}[htp]
		\centering
		\caption{Protocol Execution Time (in seconds) for Scenario I}
		\large
		\renewcommand{\arraystretch}{1.5} 
		\setlength{\tabcolsep}{6pt} 
		\resizebox{1.00\columnwidth}{!}{ 
			\begin{tabular}{|p{2.7cm}|p{5cm}|p{5cm}|p{5cm}|}
				\hline
				\textbf{Period (Month)} & \textbf{Smart Meter Execution Time ($\mathbf{t}_{sm}$)} & \textbf{Utility Provider Execution Time ($\mathbf{t}_{up}$)} & \textbf{Protocol Total Execution Time ($t_{bill} = \mathbf{t}_{sm} + \mathbf{t}_{up}$)} \\
				\hline
				\textbf{First Month} & 0.3720071145 & 0.0002259426 & 0.372233057 \\
				\hline
				\textbf{First 3 Months} & 0.9709989568 & 0.0006595637 & 0.97165852 \\
				\hline
				\textbf{First 6 Months} & 2.0005692140 & 0.0014254007 & 2.001994615 \\
				\hline
				\textbf{First 9 Months} & 2.8847361503 & 0.0020511732 & 2.886787324 \\
				\hline
				\textbf{Yearly} & 3.9425683663 & 0.0028328361 & 3.945401202 \\
				\hline
			\end{tabular}
		}
		\label{table:billing-service-evaluation}
	\end{table}
	
	\subsubsection{Memory Overhead}
	For the memory overhead, we also provide a notation table (see Table \ref{table:bill-evaluation-memory-overhead-notation}), which lists the symbols utilized in our analysis.
	
	\begin{table}[htp]
		\centering
		\caption{Notation Table for Memory Overhead}
		\normalsize
		\renewcommand{\arraystretch}{1.5}
		\setlength{\tabcolsep}{6pt}
		\resizebox{1.01\columnwidth}{!}{
			\begin{tabular}{|p{3.5cm}|p{10cm}|}
				\hline
				\textbf{Symbol} & \textbf{Description} \\
				\hline
				$\mathbf{S}_{rand}$ &
				Size of a random value generated by the secure PRNG \\
				\hline
				$\mathbf{S}_{c}$ &
				Size of a consumption value measured by the smart meter \\
				\hline
				$\mathbf{S}_{nc}$ &
				Size of a noisy consumption value \\
				\hline
				$\mathbf{S}_{trf}$ &
				Size of tariff information for one time interval \\
				\hline
				$\mathbf{S}_{s\cdot trf}$ &
				Size of a tariff-weighted noise value \\
				\hline
				$\mathbf{S}_{nb}$ &
				Size of a noisy partial bill \\
				\hline
				$\mathbf{S}_{FB}$ &
				Size of a customer's final bill \\
				\hline
			\end{tabular}
		}
		\label{table:bill-evaluation-memory-overhead-notation}
	\end{table}	
	To calculate the memory overhead at the customer premise, the smart meter retains the generated noise values required for possible tariff policy
	adjustments, the corresponding tariff information, and the tariff-weighted noise values required to compute the initial correction noise value. The
	meter also retains the final original consumption value $c_{t_L}$ for the permitted tariff policy adjustment period. Accordingly, the smart-meter
	memory overhead is bounded by
	\begin{IEEEeqnarray}{rCl}
		\mathbf{S}_{sm}
		&=&
		(L-1)\mathbf{S}_{rand}
		+
		L\mathbf{S}_{trf}
		+
		(L-1)\mathbf{S}_{s\cdot trf}
		+
		\mathbf{S}_{c}
		\label{eq:bill-protocol-evaluation-mem-overhead-sm}
	\end{IEEEeqnarray}
	The utility provider retains the noisy consumption values, noisy partial bills, and the final bill for each customer. Consequently, the memory overhead for the utility provider is as follows. 
  	\begin{IEEEeqnarray}{rCl}
		\mathbf{S}_{up}  = L (\mathbf{S}_{nc} + \mathbf{S}_{nb} + \mathbf{S}_{trf}) + \mathbf{S}_{FB} 
		\label{eq:bill-protocol-evaluation-mem-overhead-up}
	\end{IEEEeqnarray} 	
	Now, the memory overhead for the entire protocol is shown as follows. 
  	\begin{IEEEeqnarray}{rCl}
		\mathbf{S}_{bill}  = \mathbf{S}_{sm} + \mathbf{S}_{up}
		\label{eq:bill-protocol-evaluation-mem-overhead}
	\end{IEEEeqnarray}

	In our experiment, the total memory consumed by the entire protocol for a billing period (i.e., a month) is approximately 5.91484MB, which is far less than 1GB of memory dedicated for each entity.
	
	\subsubsection{Communication Overhead}
	In this work, we evaluate the communication overhead in both Neighborhood Area Networks (NAN) and Wide Area Networks (WAN) by considering the communication technologies and network protocol stacks used in each segment. In the proposed architecture, smart meters transmit their noisy consumption data to a local aggregator through IEEE 802.15.4g (Wi-SUN). Once the aggregator receives the perturbed measurements, it forwards them toward the utility provider over a 4G-LTE network. More specifically, the data is transmitted to the 4G-LTE Radio Access Network (RAN), represented by the Evolved Node B (eNB), then routed toward the Packet Data Network
	Gateway (PGW), and finally delivered to the utility provider through the corresponding network infrastructure.
	
	Table~\ref{table:network-packet-bandwidth} summarizes the estimated communication overhead for each network segment. Since each segment employs a different network protocol stack, the transmitted data size and the corresponding transmission time vary across the different segments. In this evaluation, we consider 20 smart meters and assume equal sharing of the available bandwidth among them. We assume an available bandwidth of 250~kbps for the NAN segment and 1000~kbps for each considered WAN segment. Accordingly, each smart meter is allocated 12.5~kbps in the NAN and 50~kbps in each WAN segment.
	
	The evaluation considers the main protocol payloads exchanged during billing and tariff policy adjustment, including the tariff information $\mathbf{P}_{trf}$, the adjusted tariff vector $\mathbf{P}_{TRF_{new}}$, the noisy consumption value $\mathbf{P}_{nc}$, and the generated final bill $\mathbf{P}_{FB}$. For a 30-day billing period with 15-minute reporting intervals, the adjusted tariff vector contains $L=2880$ tariff values. Assuming a 4-byte representation for each tariff value, the corresponding tariff-vector payload is therefore 11,520 bytes.
	
	\begin{table*}[htp]
		\centering
		\caption{Estimated transmitted data sizes and per-meter transmission
			times for 20 smart meters}
		\renewcommand{\arraystretch}{1.5}
		\setlength{\tabcolsep}{6pt}
		
		\resizebox{\textwidth}{!}{%
			\begin{tabular}{|>{\centering\arraybackslash}p{4.5cm}|c|c|c|c|c|c|c|c|}
				\hline
				\multirow{3}{*}{\textbf{Payload}}
				& \multicolumn{2}{c|}{\textbf{NAN}}
				& \multicolumn{6}{c|}{\textbf{WAN}} \\ \cline{2-9}
				
				& \multicolumn{2}{c|}{%
					\begin{tabular}[c]{@{}c@{}}
						\textbf{SM $\leftrightarrow$ AGG}\\
						IEEE 802.15.4g (Wi-SUN)
				\end{tabular}}
				& \multicolumn{2}{c|}{%
					\begin{tabular}[c]{@{}c@{}}
						\textbf{AGG $\leftrightarrow$ eNB}\\
						4G-LTE (PDCP-LTE)
				\end{tabular}}
				& \multicolumn{2}{c|}{%
					\begin{tabular}[c]{@{}c@{}}
						\textbf{eNB $\leftrightarrow$ PGW}\\
						IEEE 802.3 (Ethernet)
				\end{tabular}}
				& \multicolumn{2}{c|}{%
					\begin{tabular}[c]{@{}c@{}}
						\textbf{PGW $\leftrightarrow$ UP}\\
						IEEE 802.3 (Ethernet)
				\end{tabular}}
				\\ \cline{2-9}
				
				& \textbf{Transmitted Size} & \textbf{Time}
				& \textbf{Transmitted Size} & \textbf{Time}
				& \textbf{Transmitted Size} & \textbf{Time}
				& \textbf{Transmitted Size} & \textbf{Time}
				\\ \hline
				
				$\mathbf{P}_{trf}=4\,\text{B}$
				& $29\,\text{B}$ & $0.01856\,\text{s}$
				& $58\,\text{B}$ & $0.00928\,\text{s}$
				& $126\,\text{B}$ & $0.02016\,\text{s}$
				& $70\,\text{B}$ & $0.01120\,\text{s}$
				\\ \hline
				
				$\mathbf{P}_{TRF_{new}}=11{,}520\,\text{B}$
				($L=2880$)
				& $14{,}333\,\text{B}$ & $9.17312\,\text{s}$
				& $11{,}960\,\text{B}$ & $1.91360\,\text{s}$
				& $12{,}504\,\text{B}$ & $2.00064\,\text{s}$
				& $12{,}056\,\text{B}$ & $1.92896\,\text{s}$
				\\ \hline
				
				$\mathbf{P}_{nc}=4\,\text{B}$
				& $29\,\text{B}$ & $0.01856\,\text{s}$
				& $58\,\text{B}$ & $0.00928\,\text{s}$
				& $126\,\text{B}$ & $0.02016\,\text{s}$
				& $70\,\text{B}$ & $0.01120\,\text{s}$
				\\ \hline
				
				$\mathbf{P}_{FB}=4\,\text{B}$
				& $29\,\text{B}$ & $0.01856\,\text{s}$
				& $58\,\text{B}$ & $0.00928\,\text{s}$
				& $126\,\text{B}$ & $0.02016\,\text{s}$
				& $70\,\text{B}$ & $0.01120\,\text{s}$
				\\ \hline
				
				\textbf{Assumed Available BW}
				& \multicolumn{2}{c|}{250 kbps}
				& \multicolumn{2}{c|}{1000 kbps}
				& \multicolumn{2}{c|}{1000 kbps}
				& \multicolumn{2}{c|}{1000 kbps}
				\\ \hline
				
				\textbf{BW per SM}
				& \multicolumn{2}{c|}{12.5 kbps}
				& \multicolumn{2}{c|}{50 kbps}
				& \multicolumn{2}{c|}{50 kbps}
				& \multicolumn{2}{c|}{50 kbps}
				\\ \hline
			\end{tabular}%
		}
		\label{table:network-packet-bandwidth}
	\end{table*}
	
	As shown in Table~\ref{table:network-packet-bandwidth}, the regular tariff, 
	noisy consumption, and final-bill payloads are small compared with the
	adjusted tariff vector. The larger communication cost associated with a
	tariff policy adjustment mainly results from transmitting the new tariff
	vector containing all $L$ tariff values. In a Type I proportional tariff
	adjustment, no additional noisy consumption value needs to be transmitted
	by the smart meter. In contrast, each accepted Type II non-proportional
	adjustment requires only one additional final noisy consumption value from
	the smart meter after receiving the new tariff vector. Therefore, the
	proposed tariff policy adjustment mechanism introduces limited additional
	meter-to-network communication beyond the transmission of the updated tariff
	information.

	\subsection{Privacy Evaluation}
	The privacy evaluation of the proposed scheme is conducted from complementary statistical, inference-based, and distributional perspectives. First, we
	characterize the statistical properties of the proposed noise mechanism. Then, we evaluate the information exposed by the reported noisy consumption
	values under a statistical reconstruction attack and analyze the additional information introduced by tariff policy adjustments. Finally, we use the
	Jensen--Shannon (JS) divergence as a complementary metric to quantify the distributional difference between the original and perturbed consumption data.
	
	\subsubsection{Statistical Characterization of the Noise Mechanism}
	As defined in Relation~\ref{eq:bill-protocol-randomGen}, the first $L-1$
	noise values are generated independently according to
	$s_{t_i}\sim\mathcal{N}(0,\sigma^2)$. However, the final noise value is not
	independently generated. Instead, according to
	Relation~\ref{eq:bill-protocol-last-noise-generation}, it is computed as
	\begin{IEEEeqnarray}{rCl}
		s_{t_L}
		&=&
		-\sum_{i=1}^{L-1}
		\frac{trf_{t_i}}{trf_{t_L}}s_{t_i}.
		\label{eq:privacy-final-noise-linear-combination}
	\end{IEEEeqnarray}
	Since $s_{t_L}$ is a linear combination of independent zero-mean Gaussian
	random variables, its expected value is
	\begin{IEEEeqnarray}{rCl}
		\mathbb{E}[s_{t_L}]
		&=&
		-\sum_{i=1}^{L-1}
		\frac{trf_{t_i}}{trf_{t_L}}
		\mathbb{E}[s_{t_i}]
		\notag\\
		&=&0,
		\label{eq:privacy-final-noise-mean}
	\end{IEEEeqnarray}
	and its variance is
	\begin{IEEEeqnarray}{rCl}
		\operatorname{Var}(s_{t_L})
		&=&
		\sigma^2
		\sum_{i=1}^{L-1}
		\left(
		\frac{trf_{t_i}}{trf_{t_L}}
		\right)^2.
		\label{eq:privacy-final-noise-variance}
	\end{IEEEeqnarray}
	Therefore, the final correction noise follows the Gaussian distribution
	\begin{IEEEeqnarray}{rCl}
		s_{t_L}
		&\sim&
		\mathcal{N}
		\left(
		0,\,
		\sigma^2
		\sum_{i=1}^{L-1}
		\left(
		\frac{trf_{t_i}}{trf_{t_L}}
		\right)^2
		\right).
		\label{eq:privacy-final-noise-distribution}
	\end{IEEEeqnarray}
	Although the first $L-1$ noise values are independently generated, the final
	correction noise is deliberately dependent on them. More specifically, for
	$1\leq j\leq L-1$, their covariance is
	
	\begin{IEEEeqnarray}{rCl}
		\operatorname{Cov}(s_{t_j},s_{t_L})
		&=&
		-\sigma^2
		\frac{trf_{t_j}}{trf_{t_L}}.
		\label{eq:privacy-final-noise-covariance}
	\end{IEEEeqnarray}
	
	 This dependence is a direct consequence of enforcing the tariff-weighted zero-sum property required for exact billing. Under the considered threat model, the individual noise
	values are not disclosed to the semi-trusted entities; therefore, the statistical dependence alone does not directly reveal the final original
	consumption value.
	
	The implementation results are consistent with this analytical characterization. Across all 7,000 protocol runs considered in the privacy evaluation, the maximum absolute tariff-weighted zero-sum residual was $4.26326\times10^{-14}$, while the maximum absolute difference between the bill computed from the original and noisy consumption values was $2.27374\times10^{-13}$. These residuals are attributable to floating-point numerical precision and empirically confirm the exact-billing behavior of the implemented mechanism.

	\subsubsection{Reconstruction-Based Privacy Evaluation}
	Fine-grained smart-meter consumption readings exhibit temporal dependencies
	that may be exploited by semi-trusted entities together with available
	statistical background knowledge. Therefore, in addition to characterizing
	the statistical properties of the applied noise, we evaluate how much
	additional target-specific information is provided by the reported noisy
	consumption vector $NC$ under a statistical reconstruction attack.
	
	For this privacy experiment, we considered 100 eligible smart-meter traces
	and employed a meter-disjoint split consisting of 80 reference meters and
	20 held-out target meters. The reference-meter consumption profiles were
	used to construct the population-level statistical prior, while the
	consumption profiles of the 20 target meters were excluded from the
	reference set. The base noise scale $\sigma$ was computed using only the
	active-energy measurements of the reference meters. The evaluation was
	conducted over seven non-overlapping 30-day billing periods, with five
	independent noise realizations for each target meter and evaluated noise
	scale.
	
	To distinguish information that can already be inferred from population-level
	background knowledge from the additional information provided by the target
	meter's reported noisy consumption values, we consider two reconstruction
	settings. The first is a \emph{prior-only} baseline, in which the adversary
	uses the population-level statistical prior and the target customer's exact
	bill, but does not observe the target's noisy consumption vector. The second
	setting represents the evaluated reconstruction attack, in which the
	adversary is provided with the same prior information and exact bill together
	with the target's reported noisy consumption vector $NC$. The adversary then
	applies a temporal linear minimum mean-square error (LMMSE) estimator and
	projects the resulting estimate onto the exact-billing constraint. Hence,
	the difference between the two reconstruction settings isolates the
	additional reconstruction capability provided by access to the target's
	noisy reports.
	
	Let $\hat{C}$ denote the reconstructed consumption profile. We first measure
	the reconstruction error using the root mean-square error (RMSE):
	\begin{IEEEeqnarray}{rCl}
		RMSE(C,\hat{C})
		&=&
		\sqrt{
			\dfrac{1}{L}
			\sum_{i=1}^{L}
			(c_{t_i}-\hat{c}_{t_i})^2
		}.
		\label{eq:privacy-reconstruction-rmse}
	\end{IEEEeqnarray}
	
	We also compute the Pearson correlation coefficient between the original and
	reconstructed consumption profiles as a complementary measure of similarity
	in their temporal patterns. A lower RMSE indicates more accurate
	reconstruction, while a higher correlation indicates that the reconstructed
	profile more closely follows the temporal variation of the original
	consumption data.
	
	To quantify the additional reconstruction capability obtained specifically
	from the target's noisy reports, let $RMSE_{\mathrm{prior}}$ denote the RMSE
	of the prior-only estimator and $RMSE_{\mathrm{attack}}$ denote the RMSE of
	the estimator that additionally observes $NC$. We define the incremental
	reconstruction advantage as
	
	\begin{IEEEeqnarray}{rCl}
		Adv_{\mathrm{recon}}
		&=&
		100
		\left(
		\dfrac{
			RMSE_{\mathrm{prior}}-RMSE_{\mathrm{attack}}
		}{
			RMSE_{\mathrm{prior}}
		}
		\right).
		\label{eq:privacy-reconstruction-advantage}
	\end{IEEEeqnarray}
	
	Accordingly, a larger positive value of $Adv_{\mathrm{recon}}$ indicates
	that the target's noisy consumption reports provide substantial additional
	information beyond the population prior, whereas a value close to zero
	indicates that access to $NC$ provides little additional improvement in
	reconstruction accuracy.
	
\begin{center}
		\begin{figure}[htp]
			\includegraphics[width=3.5in]{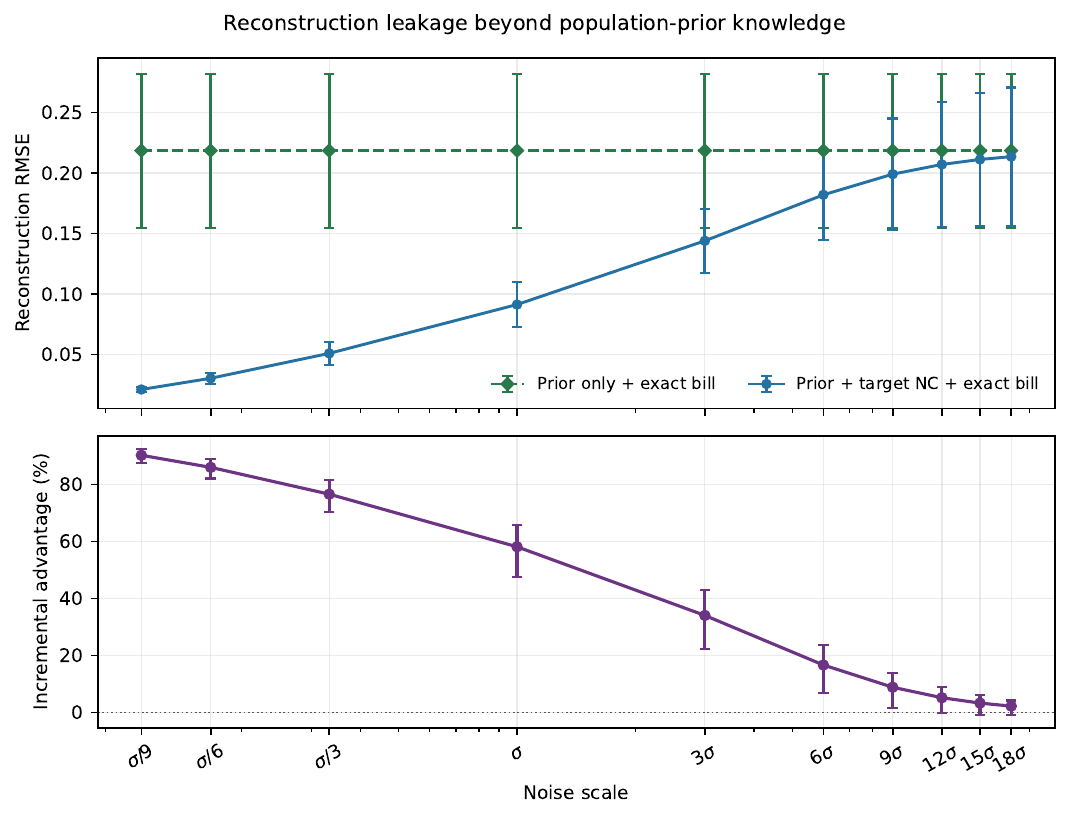}
			\caption{Reconstruction leakage beyond population-prior knowledge.
				The upper panel compares the prior-only estimator with the statistical
				reconstruction attack that additionally receives the target's noisy
				consumption reports, while both estimators use the exact bill. The
				lower panel shows the incremental RMSE advantage attributable to the
				target's noisy reports.}
			\label{fig:privacy-reconstruction-leakage}
		\end{figure}
	\end{center}

	Fig.~\ref{fig:privacy-reconstruction-leakage} presents the aggregated
	reconstruction results across the evaluated noise scales. At the base noise
	scale $\sigma$, access to the target's noisy consumption reports provides an
	incremental reconstruction advantage of $58.19\%$ over the prior-only
	baseline. As the noise scale increases, this additional reconstruction
	capability decreases considerably, reaching $8.90\%$ at $9\sigma$ and
	$2.28\%$ at the highest evaluated stress-test scale of $18\sigma$.
	Accordingly, the reconstruction error of the evaluated attack progressively
	approaches that of the prior-only baseline as the perturbation level
	increases.
	
	At $18\sigma$, the $95\%$ confidence interval of the incremental
	reconstruction advantage is $[-0.733\%,\,4.247\%]$, which includes zero.
	Therefore, under the considered statistical reconstruction attack, the noisy
	consumption reports do not provide a statistically reliable improvement over
	the prior-only reconstruction at this stress-test scale.
	
\begin{center}
		\begin{figure}[htp]
			\includegraphics[width=3.5in]{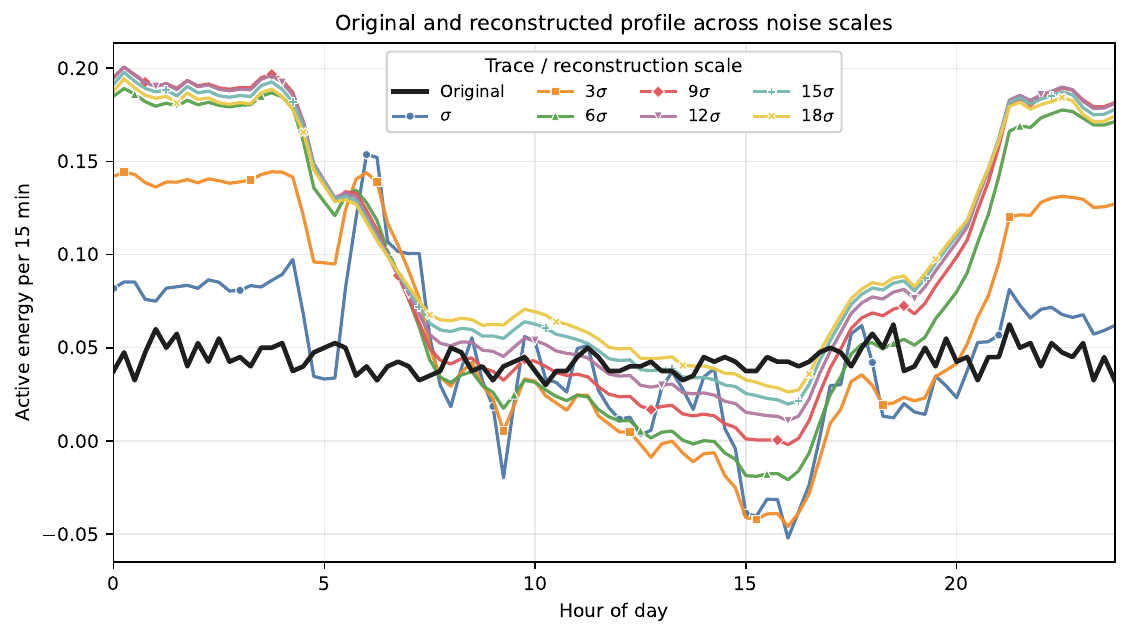}
			\caption{Illustrative reconstruction of a preselected held-out daily
				consumption profile across representative noise scales.}
			\label{fig:privacy-profile-reconstruction-scales}
		\end{figure}
	\end{center}

	For additional qualitative insight,
	Fig.~\ref{fig:privacy-profile-reconstruction-scales} illustrates the
	reconstruction of a preselected held-out daily consumption profile across
	representative noise scales. The target meter, billing period, and noise
	realization were fixed before examining the reconstruction outcomes. As the
	noise scale increases, the reconstructed profile progressively loses
	target-specific variations and becomes more influenced by the population-level
	prior. This qualitative example complements the aggregated reconstruction
	results presented in Fig.~\ref{fig:privacy-reconstruction-leakage}.
	
	\subsubsection{Privacy Analysis Under Tariff Policy Adjustments}
	
	We further analyze whether tariff policy adjustments introduce additional
	information about the customer's original consumption profile. Since the
	proposed mechanism preserves exact billing, each final bill represents a
	tariff-weighted relation among the $L$ original consumption values.
	
	For a Type I proportional adjustment, the new tariff vector is defined as
	\begin{IEEEeqnarray}{rCl}
		TRF_{new}=\alpha TRF,
		\qquad \alpha>0.
		\label{eq:privacy-type-I-proportional}
	\end{IEEEeqnarray}
	Consequently, the adjusted billing relation is only a scaled version of the
	original one and does not provide an additional independent equation about
	the consumption profile. Moreover, as discussed earlier, a Type I adjustment
	requires no additional noisy consumption report and does not increment $K$.
	
	In contrast, a Type II adjustment changes the tariff vector
	non-proportionally. For the $k$-th accepted Type II adjustment, the
	corresponding exact bill can be expressed as
	\begin{IEEEeqnarray}{rCl}
		FB_{id_j}^{(k)}
		&=&
		\sum_{i=1}^{L}
		c_{t_i}trf_{t_i}^{(k)}.
		\label{eq:privacy-type-II-billing-relation}
	\end{IEEEeqnarray}
	Therefore, each accepted Type II adjustment may introduce at most one
	additional independent linear equation involving the same $L$ original
	consumption values. Considering the original billing relation together with
	$K$ accepted Type II adjustments, the number of linearly independent billing
	equations, denoted by $q$, satisfies
	
	\begin{IEEEeqnarray}{rCl}
		q \leq K+1.
		\label{eq:privacy-type-II-rank-bound}
	\end{IEEEeqnarray}
	
	Accordingly, the number of remaining unconstrained dimensions of the
	consumption profile satisfies
	\begin{IEEEeqnarray}{rCl}
		L-q
		&\geq&
		L-(K+1).
		\label{eq:privacy-type-II-degrees-freedom}
	\end{IEEEeqnarray}
	
	Since the protocol restricts the number of accepted non-proportional
	adjustments by $K\leq K_{\max}$, the additional exact information introduced
	through repeated Type II adjustments is correspondingly bounded. This
	analysis characterizes the algebraic information provided by the billing
	relations alone; statistical inference based on noisy consumption reports
	and background knowledge is evaluated separately in the reconstruction
	analysis.

	\subsubsection{Distributional Privacy Evaluation Using Jensen--Shannon Divergence}
	
	In addition to the reconstruction-based evaluation, we employ the
	Jensen--Shannon (JS) divergence to quantify the distributional difference
	between the original and noisy consumption data. Let $P$ and $Q$ denote the
	empirical probability distributions of the original and perturbed consumption
	values, respectively, and let
	\begin{IEEEeqnarray}{rCl}
		M
		&=&
		\dfrac{1}{2}(P+Q).
		\label{eq:privacy-js-mixture}
	\end{IEEEeqnarray}
	The JS divergence is then defined as
	\begin{IEEEeqnarray}{rCl}
		D_{JS}(P\|Q)
		&=&
		\dfrac{1}{2}D_{KL}(P\|M)
		+
		\dfrac{1}{2}D_{KL}(Q\|M).
		\label{eq:privacy-js-divergence}
	\end{IEEEeqnarray}
	Using base-2 logarithms, the JS divergence is bounded between 0 and 1,
	where a larger value indicates a greater distributional difference between
	the original and perturbed consumption data.
	
\begin{center}
		\begin{figure}[htp]
			\includegraphics[width=3.5in]{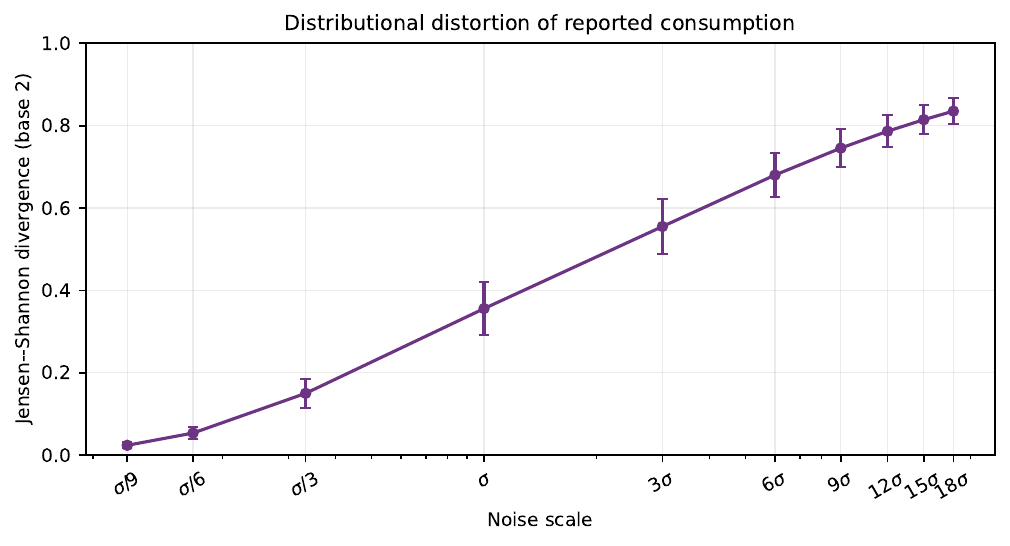}
			\caption{Distributional distortion between the original and noisy
				consumption data measured using base-2 Jensen--Shannon divergence
				across the evaluated noise scales.}
			\label{fig:privacy-js-divergence}
		\end{figure}
	\end{center}

	For a consistent comparison across the evaluated noise scales, the same
	histogram boundaries were used for both the original and noisy consumption
	distributions. Fig.~\ref{fig:privacy-js-divergence} shows that the mean JS
	divergence increases progressively as the applied noise scale increases. In
	particular, the mean divergence increases from $0.356$ at $\sigma$ to
	$0.746$ at $9\sigma$ and $0.835$ at the $18\sigma$ stress-test scale. The
	same increasing trend was also observed when the analysis was repeated using
	50, 100, and 200 histogram bins.
	
	Using base-2 logarithms, the JS divergence is bounded between 0 and 1.
	Within this distributional evaluation, values closer to 1 indicate a
	greater separation between the original and perturbed consumption
	distributions and therefore stronger distributional obfuscation, whereas
	values closer to 0 indicate greater similarity between the two distributions. 
	
	The JS divergence therefore provides complementary distributional evidence to the reconstruction-based analysis and should not be interpreted as a standalone formal privacy guarantee.

	\section{Conclusion}
	In this paper, we proposed a lightweight, privacy-preserving smart metering billing protocol based on real-time tariffs with dynamic tariff policy adjustment. Smart meters report their fine-grained consumption values to the utility provider for various services, such as accurate dynamic billing. Subsequently, the utility provider generates weekly or monthly bills for customers based on the policies adopted by the energy company.
	
	While such fine-grained readings are essential for billing, operational, and value-added services, they can also reveal sensitive private information, including users' presence or absence, daily routines, and behavioral patterns. To address these privacy concerns, we introduced a secure and privacy-aware billing system that supports real-time pricing policies and dynamic tariff policy adjustments. Our scheme leverages a lightweight data perturbation mechanism with a tariff-weighted zero-sum property to protect fine-grained consumption readings while preserving exact billing.
	
	In the first scenario, smart meters report noisy consumption values to the utility provider at each time interval. The utility provider applies the corresponding tariffs to these values and computes the customer's final bill. In the second scenario, the utility provider can recompute customers' bills following an authorized tariff policy adjustment. For proportional Type I adjustments, the previously reported noisy consumption values can be reused without any additional report from the smart meter. For non-proportional Type II adjustments, the smart meter recomputes only the final correction noise value and reports the updated final noisy consumption value, while the previous $L-1$ noisy readings are reused.
	
	Furthermore, we conducted analytical and experimental evaluations of the proposed scheme in terms of computational, memory, communication, and privacy characteristics. According to our experimental outcomes, the protocol's execution time for a complete year is approximately $3.94540$ seconds, demonstrating its lightweight computational requirements. The experimental results also confirmed that the tariff-weighted correction mechanism preserves exact billing to numerical precision.
	
	Lastly, we evaluated privacy from complementary statistical, reconstruction-based, and distributional perspectives. The reconstruction results showed that increasing the noise scale reduces the additional information that can be obtained from the reported noisy consumption values, while the Jensen--Shannon divergence results showed an increasing distributional difference between the original and perturbed consumption data. These results show that increasing the noise scale can strengthen the privacy protection provided by the perturbation mechanism under the evaluated attacks, while preserving exact billing and lightweight
	operation.

	
	
	%
	
	



	\ifCLASSOPTIONcaptionsoff
	\newpage
	\fi

	
	
	
	
	\bibliographystyle{IEEEtran}
	\bibliography{IEEEabrv,Bibliography}

@IEEEtranBSTCTL{IEEEexample:BSTcontrol,
    CTLuse_article_number = "yes",
    CTLuse_paper = "yes",
    CTLuse_forced_etal = "no",
    CTLmax_names_forced_etal = "50",
    CTLnames_show_etal = "50",
    CTLuse_alt_spacing = "yes",
    CTLalt_stretch_factor = "4",
    CTLdash_repeated_names = "yes",
    CTLname_format_string = "{f.~}{vv~}{ll}{, jj}",
    CTLname_latex_cmd = "",
    CTLname_url_prefix = "[Online]. Available:"
 }

@article{kumar2019smart,
  title={Smart grid metering networks: A survey on security, privacy and open research issues},
  author={Kumar, Pardeep and Lin, Yun and Bai, Guangdong and Paverd, Andrew and Dong, Jin Song and Martin, Andrew},
  journal={IEEE Communications Surveys \& Tutorials},
  volume={21},
  number={3},
  pages={2886--2927},
  year={2019},
  publisher={IEEE}
}

@article{souri2014smart,
  title={Smart metering privacy-preserving techniques in a nutshell},
  author={Souri, Hajer and Dhraief, Amine and Tlili, Syrine and Drira, Khalil and Belghith, Abdelfettah},
  journal={Procedia Computer Science},
  volume={32},
  pages={1087--1094},
  year={2014},
  publisher={Elsevier}
}

@article{ansari2022state,
  title={State of art and comprehensive study on smart meter networking},
  author={Ansari, Aftab Ahmed and Giribabu, Dyanamina},
  journal={Flexible Electronics for Electric Vehicles: Select Proceedings of FlexEV—2021},
  pages={377--393},
  year={2022},
  publisher={Springer}
}

@article{asghar2017smart,
  title={Smart meter data privacy: A survey},
  author={Asghar, Muhammad Rizwan and D{\'a}n, Gy{\"o}rgy and Miorandi, Daniele and Chlamtac, Imrich},
  journal={IEEE Communications Surveys \& Tutorials},
  volume={19},
  number={4},
  pages={2820--2835},
  year={2017},
  publisher={IEEE}
}

@inproceedings{kayalvizhy2021survey,
  title={A survey on cyber security attacks and countermeasures in smart grid metering network},
  author={Kayalvizhy, V and Banumathi, A},
  booktitle={2021 5th International Conference on Computing Methodologies and Communication (ICCMC)},
  pages={160--165},
  year={2021},
  organization={IEEE}
}

@article{kua2023privacy,
  title={Privacy preservation in smart meters: current status, challenges and future directions},
  author={Kua, Jonathan and Hossain, Mohammad Belayet and Natgunanathan, Iynkaran and Xiang, Yong},
  journal={Sensors},
  volume={23},
  number={7},
  pages={3697},
  year={2023},
  publisher={MDPI}
}

@inproceedings{siddiqui2012smart,
  title={Smart grid privacy: Issues and solutions},
  author={Siddiqui, Farhan and Zeadally, Sherali and Alcaraz, Cristina and Galvao, Samara},
  booktitle={2012 21st International Conference on Computer Communications and Networks (ICCCN)},
  pages={1--5},
  year={2012},
  organization={IEEE}
}

@inproceedings{bohli2010privacy,
  title={A privacy model for smart metering},
  author={Bohli, Jens-Matthias and Sorge, Christoph and Ugus, Osman},
  booktitle={2010 IEEE International Conference on Communications Workshops},
  pages={1--5},
  year={2010},
  organization={IEEE}
}

@article{wang2023privacy,
  title={Privacy-preserving data aggregation with dynamic billing in fog-based smart grid},
  author={Wang, Huiyong and Gong, Yunmei and Ding, Yong and Tang, Shijie and Wang, Yujue},
  journal={Applied Sciences},
  volume={13},
  number={2},
  pages={748},
  year={2023},
  publisher={MDPI}
}

@inproceedings{jawurek2011plug,
  title={Plug-in privacy for smart metering billing},
  author={Jawurek, Marek and Johns, Martin and Kerschbaum, Florian},
  booktitle={International Symposium on Privacy Enhancing Technologies Symposium},
  pages={192--210},
  year={2011},
  organization={Springer}
}

@inproceedings{molina2010private,
  title={Private memoirs of a smart meter},
  author={Molina-Markham, Andr{\'e}s and Shenoy, Prashant and Fu, Kevin and Cecchet, Emmanuel and Irwin, David},
  booktitle={Proceedings of the 2nd ACM workshop on embedded sensing systems for energy-efficiency in building},
  pages={61--66},
  year={2010}
}

@inproceedings{rial2011privacy,
  title={Privacy-preserving smart metering},
  author={Rial, Alfredo and Danezis, George},
  booktitle={Proceedings of the 10th annual ACM workshop on Privacy in the electronic society},
  pages={49--60},
  year={2011}
}

@article{rial2018privacy,
  title={Privacy-preserving smart metering revisited},
  author={Rial, Alfredo and Danezis, George and Kohlweiss, Markulf},
  journal={International Journal of Information Security},
  volume={17},
  pages={1--31},
  year={2018},
  publisher={Springer}
}

@inproceedings{groth2005non,
  title={Non-interactive zero-knowledge arguments for voting},
  author={Groth, Jens},
  booktitle={Applied Cryptography and Network Security: Third International Conference, ACNS 2005, New York, NY, USA, June 7-10, 2005. Proceedings 3},
  pages={467--482},
  year={2005},
  organization={Springer}
}

@article{eccles2017performance,
  title={Performance analysis of secure and private billing protocols for smart metering},
  author={Eccles, Tom and Halak, Basel},
  journal={Cryptography},
  volume={1},
  number={3},
  pages={20},
  year={2017},
  publisher={MDPI}
}

@article{petrlic2010privacy,
  title={A privacy-preserving concept for smart grids},
  author={Petrlic, Ronald},
  journal={Sicherheit in vernetzten Systemen},
  volume={18},
  pages={B1--B14},
  year={2010}
}

@inproceedings{ben2021efficient,
  title={An efficient and privacy-preserving billing protocol for smart metering},
  author={Ben Romdhane, Rihem and Hammami, Hamza and Hamdi, Mohamed and Kim, Tai-Hoon},
  booktitle={International Conference on Advanced Information Networking and Applications},
  pages={691--702},
  year={2021},
  organization={Springer}
}

@article{zhang2020privacy,
  title={Privacy-Functionality Trade-Off: A privacy-preserving multi-channel smart metering system},
  author={Zhang, Xiao-Yu and Kuenzel, Stefanie and C{\'o}rdoba-Pach{\'o}n, Jos{\'e}-Rodrigo and Watkins, Chris},
  journal={Energies},
  volume={13},
  number={12},
  pages={3221},
  year={2020},
  publisher={MDPI}
}

@inproceedings{efthymiou2010smart,
  title={Smart grid privacy via anonymization of smart metering data},
  author={Efthymiou, Costas and Kalogridis, Georgios},
  booktitle={2010 first IEEE international conference on smart grid communications},
  pages={238--243},
  year={2010},
  organization={IEEE}
}

@inproceedings{ababneh2022private,
  title={Private and Secure Smart Meter Billing},
  author={Ababneh, Mohammed and Kolachala, Kartick and Vishwanathan, Roopa},
  booktitle={Proceedings of the 8th ACM on Cyber-Physical System Security Workshop},
  pages={15--25},
  year={2022}
}

@article{li2023fine,
  title={A fine-grained privacy protection data aggregation scheme for outsourcing smart grid},
  author={Li, Hongyang and Li, Xinghua and Cheng, Qingfeng},
  journal={Frontiers of Computer Science},
  volume={17},
  number={3},
  pages={173806},
  year={2023},
  publisher={Springer}
}

@article{xu2023privacy,
  title={A privacy-preserving framework using homomorphic encryption for smart metering systems},
  author={Xu, Weiyan and Sun, Jack and Cardell-Oliver, Rachel and Mian, Ajmal and Hong, Jin B},
  journal={Sensors},
  volume={23},
  number={10},
  pages={4746},
  year={2023},
  publisher={MDPI}
}

@article{abidin2018secure,
  title={Secure and privacy-friendly local electricity trading and billing in smart grid},
  author={Abidin, Aysajan and Aly, Abdelrahaman and Cleemput, Sara and Mustafa, Mustafa A},
  journal={arXiv preprint arXiv:1801.08354},
  year={2018}
}

@article{nabil2019ppetd,
  title={PPETD: Privacy-preserving electricity theft detection scheme with load monitoring and billing for AMI networks},
  author={Nabil, Mahmoud and Ismail, Muhammad and Mahmoud, Mohamed MEA and Alasmary, Waleed and Serpedin, Erchin},
  journal={Ieee Access},
  volume={7},
  pages={96334--96348},
  year={2019},
  publisher={IEEE}
}

@article{sultan2019privacy,
  title={Privacy-preserving metering in smart grid for billing, operational metering, and incentive-based schemes: A survey},
  author={Sultan, Sari},
  journal={Computers \& Security},
  volume={84},
  pages={148--165},
  year={2019},
  publisher={Elsevier}
}

@article{cenky2023dataset,
  title={Dataset of 15-minute values of active and reactive power consumption of 1000 households during single year},
  author={Cenk{\`y}, Matej and Bend{\'\i}k, Jozef and Cintula, Boris and Janiga, Peter and Eleschov{\'a}, {\v{Z}}aneta and Bel{\'a}{\v{n}}, Anton},
  journal={Data in Brief},
  volume={50},
  pages={109588},
  year={2023},
  publisher={Elsevier}
}

@techreport{DOE2020SmartGrid,
  author = {{U.S. Department of Energy}},
  title = {2020 Smart Grid System Report},
  institution = {United States Department of Energy},
  year = {2022},
  month = {May},
  url = {https://www.energy.gov/sites/default/files/2022-05/2020%20Smart%20Grid%20System%20Report_0.pdf},
}

@misc{EuropeanCommission2025,
    author        = "{European Commission}",
    title         = "Smart grids and meters",
    year          = {2025},
    url           = "https://energy.ec.europa.eu/topics/markets-and-consumers/smart-grids-and-meters_en",
    note          = "[Accessed: March 8, 2025]"
}

@misc{MarketsAndData2025,
    author        = "{Markets and Data}",
    title         = "United States Smart Grid Market Assessment, Opportunities, and Forecast, 2016-2030",
    year          = {2025},
    url           = "https://www.marketsandata.com/industry-reports/united-states-smart-grid-market",
    note          = "[Accessed: March 8, 2025]"
}

@misc{IMARC2025,
    author        = "{IMARC Group}",
    title         = "UK Smart Grid Market: Industry Trends, Share, Size, Growth, Opportunity and Forecast",
    year          = {2025},
    url           = "https://www.imarcgroup.com/uk-smart-grid-market",
    note          = "[Accessed: March 8, 2025]"
}

@article{oksuz2024consortium,
  author={Oksuz, Ozgur},
  title={Consortium blockchain based secure and efficient data aggregation and dynamic billing system in smart grid},
  journal={Peer-to-Peer Networking and Applications},
  volume={17},
  pages={2717--2736},
  year={2024},
  doi={10.1007/s12083-024-01709-8}
}

@inproceedings{kayumov2025trusted,
  author={Kayumov, Abduaziz and Lee, Shincheol and Shin, Ji Sun},
  title={Trusted and Privacy-Preserving Billing System for Smart Grids},
  booktitle={2025 IEEE International Conference on Big Data and Smart Computing (BigComp)},
  pages={160--166},
  year={2025},
  publisher={IEEE},
  doi={10.1109/BigComp64353.2025.00043}
}

@article{wagh2024distributed,
  author={Wagh, Gaurav S. and Mishra, Sumita and Agarwal, Anurag},
  title={A Distributed Privacy-Preserving Framework With Integrity Verification Capabilities for Metering Data and Dynamic Billing in a Malicious Setting},
  journal={IEEE Transactions on Smart Grid},
  volume={15},
  number={6},
  pages={5813--5825},
  year={2024},
  doi={10.1109/TSG.2024.3433385}
}

@article{braeken2018efficient,
  author={Braeken, An and Kumar, Pardeep and Martin, Andrew},
  title={Efficient and Privacy-Preserving Data Aggregation and Dynamic Billing in Smart Grid Metering Networks},
  journal={Energies},
  volume={11},
  number={8},
  pages={2085},
  year={2018},
  doi={10.3390/en11082085}
}

@article{ibrahem2021efficient,
  author={Ibrahem, Mohamed I. and Nabil, Mahmoud and Fouda, Mostafa M. and Mahmoud, Mohamed M. E. A. and Alasmary, Waleed and Alsolami, Fawaz},
  title={Efficient Privacy-Preserving Electricity Theft Detection With Dynamic Billing and Load Monitoring for AMI Networks},
  journal={IEEE Internet of Things Journal},
  volume={8},
  number={2},
  pages={1243--1258},
  year={2021},
  doi={10.1109/JIOT.2020.3026692}
}

	\vfill
	

\end{document}